\documentclass{article}
\usepackage{arxiv}
\usepackage[utf8]{inputenc} 
\usepackage[T1]{fontenc}    
\usepackage{hyperref}       
\usepackage{url}            
\usepackage{booktabs}       
\usepackage{amsfonts}       
\usepackage{nicefrac}       
\usepackage{microtype}      
\usepackage{lipsum}		
\usepackage{natbib}
\usepackage{doi}
\usepackage{psfrag,color}
\usepackage[dvips]{graphicx}
\graphicspath{{./Figures/}} 
\DeclareGraphicsExtensions{.eps} 

\title{A Markov Decision Process Model for Intrusion Tolerance Problems}

\author{{O.~Patrick Kreidl}\thanks{This manuscript is based upon the 2010 submission to {\it IEEE Transactions on Dependable and Secure Computing}, which was rejected} \\
	School of Engineering (with Courtesy Appointment in School of Computing)\\
	University of North Florida\\
	Jacksonville, FL 12224 \\
	\texttt{patrick.kreidl@unf.edu} \\
}

\date{}

\renewcommand{\headeright}{O.~Patrick Kreidl}
\renewcommand{\undertitle}{}
\renewcommand{\shorttitle}{A Markov Decision Process Model for Intrusion Tolerance Problems}

\newcommand{\comment}[1]{}
\newcommand{\sSecRef}[1]{Subsection~\ref{ssec:#1}}
\newcommand{\equRef}[1]{(\ref{equ:#1})}
\newcommand{\equRefs}[2]{(\ref{equ:#1})-(\ref{equ:#2})}
\newcommand{\EquRefs}[2]{Equations~(\ref{equ:#1})-(\ref{equ:#2})}
\newcommand{\allZ}{\ensuremath{\mathbb{Z}}}
\newcommand{\allU}{\ensuremath{\mathbb{U}}}
\newcommand{\allX}{\ensuremath{\mathbb{X}}}
\newcommand{\allT}{\ensuremath{\mathbb{T}}}
\newcommand{\mb}[1]{\ensuremath{\mathbf{#1}}}

\nocite{*}

\begin{document}
\maketitle

\begin{abstract}
  We formulate and analyze a simplest Markov decision process model
  for intrusion tolerance problems, assuming that (i)~each attack
  proceeds through one or more steps before the system's security
  fails, (ii)~defensive responses that target these intermediate steps
  may only sometimes thwart the attack and (iii)~reset responses that
  are sensible upon discovering an attack's completion may not always
  recover from the security failure. The analysis shows that, even in
  the ideal case of perfect detectors, it can be sub-optimal in the
  long run to employ defensive responses while under attack; that is,
  depending on attack dynamics and response effectiveness, the total
  overhead of ongoing defensive countermeasures can exceed the total
  risk of intermittent security failures. The analysis similarly
  examines the availability loss versus the risk reduction of
  employing preemptive resets, isolating key factors that determine
  whether system recovery is best initiated reactively or proactively.
  We also discuss model extensions and related work looking towards
  intrusion tolerance applications with (i)~imperfect or controllable 
  detectors, (ii)~multiple types of attacks, (iii)~continuous-time
  dynamics or (iv)~strategic attackers.
\end{abstract}

\newpage
\section*{Ackowledgments}
\begin{quote}
``If at first you don't succeed, give up.'' \quad {\it ---Homer Simpson.}
\end{quote}

This manuscript is based upon the 2010 submission to {\it IEEE Transactions on Dependable and Secure Computing}, which was rejected. The reviewers deemed the main model as overly-simplified and, in turn, the implications of the analysis of little practical relevance. There was also disappointment that the model's various extensions were only summarized, advising to form future submissions out of thorough analyses of each extension. The reviewers' negative feedback was, interestingly, opposing the positive feedback a preliminary version of this work received earlier in the year at the \emph{2010 Workshop on Recent Advances in Intrusion-Tolerant Systems (WRAITS)}, which preceded the {\it 2010 IEEE/IFIP International Conference on Dependable Systems and Networks (DSN)} in Chicago, IL. The workshop participants appreciated the simplified abstraction in the proposed model and found value in the conclusions, especially around the conditions under which defensive responses are worthwhile or not. This preliminary model made the strong assumption that the ``Reset'' action was perfect and returned the system to ``Normal'' (i.e., returning to state $N$) immediately upon a security faliure (i.e., upon entering state $F$). That is, in terms of the model in this manuscript, the WRAITS work considered only the special case of choosing probablity parameter $p_R = 1$. It was another workshop speaker, Professor Robert Constable from Cornell University, who questioned the assumption, inspiring this manuscript to generalize the analysis for cases in which parameter $p_R < 1$. Due to the intelluctual influece of WRAITS on this manuscript, the workshop's organizers and program committee are gratefully acknowledged: Saurabh Bagchi (Purdue University, USA), Byung-Gon Chun (Intel Labs Berkeley, USA), Manuel Costa (Microsoft Research, UK), Flavio Junqueira (Yahoo! Research, Spain), Rama Kotla (Microsoft Research, USA), Peng Liu (Penn State University, USA), Jean-Phillipe Martin (Microsoft Research, UK), Nuno Neves (University of Lisboa, Portugal), Rodrigo Rodrigues (MPI-SWS, Germany), William H. Sanders (University of Illinois UC, USA), Arun Sood (George Mason University, USA) and Paulo Verissimo (University of Lisboa, Portugal).

Today, it is \today ~and the author keeps no ambition to conventionally publish this manuscript, so is choosing to upload it to the \emph{arXiv} open-access online repository. The manuscript is essentially unreviewed, and the only addition to its 2010 version is on the model extension of strategic attackers. Readers are invited to communicate questions or errata to the author via email: {\tt patrick.kreidl@unf.edu}

\setlength{\topmargin}{-0.5in}
\setlength{\headsep}{0.5in}
\pagestyle{myheadings}
\addtocounter{page}{-1}
\markright{A Markov Decision Process Model for Intrusion Tolerance Problems \hfill 
Kreidl--}



\newcommand{\FigRef}[1]{Fig.~\ref{fig:#1}}
\newcommand{\SecRef}[1]{Section~\ref{sec:#1}}

\newpage
\section{Introduction}
\begin{figure}[!h]
\centering
\psfrag{Information}[c][c]{\small Information}
\psfrag{System}[c][c]{\small System}
\psfrag{Sensing}[c][c]{\small Sensing}
\psfrag{Devices}[c][c]{\small Devices}
\psfrag{Response}[c][c]{\small Response}
\psfrag{Security}[c][c]{\small Security}
\psfrag{Controller}[c][c]{\small Controller}
\psfrag{observations}[c][c]{\small \it observations}
\psfrag{commands}[c][c]{\small \it commands}
\resizebox{!}{1.25in}{\includegraphics{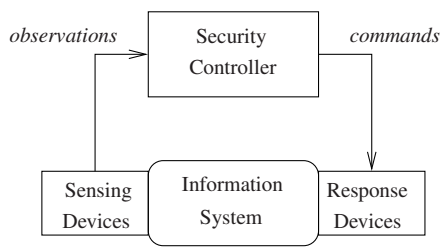}} \\
\caption{ \label{fig:systemDiagram} An intrusion tolerant system.}
\end{figure}

Intrusion tolerance is an active area of research
that aims to extend proven solutions in the fields of fault-tolerance
and dependability to address growing concerns about weaknesses of
modern-day information security systems \cite{DBF91:InTol,VNC03:InTol}. The key
assumption is that, even when state-of-the-art intrusion prevention
and detection techniques are in place, a security breach will occur
eventually.  An intrusion tolerant system (see \FigRef{systemDiagram})
is therefore also equipped with mechanisms for reacting in real-time
to an ongoing intrusion, striving not just to better detect its
occurrence but also to diagnose or predict its evolution so that
defensive countermeasures can be invoked timely enough to delay,
frustrate or altogether thwart the attacker's ultimate
intentions. Successful realization of this vision introduces numerous
technical challenges, including the embedded design of real-time
sensing/response devices as well as the automatic control of these
devices to favorably influence the system's security status over its
operational lifetime.

This paper analyzes the problem of intrusion tolerance within the
formalism of Markov Decision Processes (MDPs) \cite{Ber95:DPOC1}. MDPs are
related to the types of stochastic models already popular in
contemporary dependability/security research \cite{JoO97:QuMod, SCW02:ModEV,
  MGV02:MetMQ, ASH05:RTIDS, SHK06:GaThe}.  Their focus is to assume that the model
parameters (e.g., state transition probabilities \& costs) in each
decision stage depend explicitly on the control action taken in that
stage. For example, an MDP can capture the reinstatement of security
(and the possible cost of disrupting all normal services, or loss of
availability) upon taking a cleansing "reset" action (e.g., initiating
a server rotation \cite{HAS06:SCITb} or perhaps a more elaborate recovery
sequence \cite{JSH06:AutRe}); as another example, an MDP can capture the
possible prevention of a security failure from a well-timed "defend"
action (e.g., killing a process that is critical to the intrusion's
continuation \cite{KrF04:aLADS}). In any case, the expected performance
(e.g., mean-time-to-failure) is dependent on the \emph{policy}, or
rule that dictates how control actions are selected, and the
\emph{optimal policy} is one that achieves the best performance for a
given model.

\SecRef{MainModel} presents a simplest MDP model for intrusion
tolerance problems. Though the model assumes only three system states
and only three candidate controls for ease of analysis, it is
sufficiently rich to expose a number of important policy
considerations for intrusion tolerant systems. Even in the ideal case
of perfect detectors (\SecRef{PerDet}), it is sometimes sub-optimal in
the long run to employ a policy that invokes defensive responses
during the intermediate stages of an intrusion; more specifically,
depending on attack dynamics and response effectiveness, policies
involving only occasional reset actions may provide a better tradeoff
between the total overhead of ongoing security maintenance and the
total risk of potential security failures. The model also includes the
possibility of employing reset actions preemptively, capturing in fundamental
terms the long-appreciated tradeoff between availability loss and risk reduction
\cite{ZSR03:FoITS,SNV06:ProRe} as well as isolating key factors that determine whether system
recovery is best initiated reactively or proactively \cite{SBC07:ResIT}; interestingly,
whether a security failure is equated with catastrophic consequences is
just one of several factors, another being e.g., whether the reset action is
sufficiently reliable in comparison to attack dynamics (frequency, speed, etc.).
%
Possible model extensions (and related work) are suggested in
\SecRef{ModExt}, looking towards intrusion tolerance applications with
(i)~imperfect or controllable detectors, (ii)~multiple types of
attacks, (iii)~continuous-time dynamics or (iv)~strategic
attackers. We conclude in \SecRef{Conclu}.

\section{Main Model\label{sec:MainModel}}
This section abstracts the problem of intrusion tolerance within the
modeling formalism of a stationary, discrete-time, finite-state,
finite-action Markov Decision Process (MDP) \cite{Ber95:DPOC1}. Underlying
such a model is (i)~a finite collection of finite-state Markov chains,
one per candidate control action, and (ii)~the assumption that
decisions are made in discrete stages: each stage begins in a
particular state and, upon taking a control action, the process
probabilistically transitions to a new state (or self-transitions to
the same state), simultaneously ending the current stage and beginning
the next.  Thus, the outcome of each decision stage is defined by the
selected control and the realized state transition. In turn, an MDP
model associates both a probability and a cost to every such outcome.

\begin{figure}[!t]
\centering
\psfrag{N}[c][c]{$N$}
\psfrag{A}[c][c]{$A$}
\psfrag{F}[c][c]{$F$}
\psfrag{pA}[c][c]{\small $p_A,0$}
\psfrag{qA}[l][l]{\small $\!\!1-p_A,0$}
\psfrag{pF}[c][c]{\small $p_F,c_A$}
\psfrag{qF}[l][l]{\small $1-p_F,c_A$}
\psfrag{1}[c][l]{\small $1,c_F$}
\includegraphics{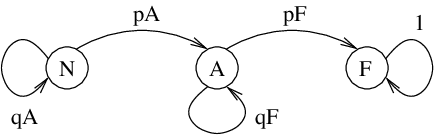} \\
{\small (a) Under a ``Wait'' action (Control $W$)}\\[0.3in]
\psfrag{pA}[c][c]{\small $p_A,c_D$}
\psfrag{qA}[l][l]{\small $\!\!1-p_A,c_D$}
\psfrag{pD}[t][t]{\small $p_D,c_D$}
\psfrag{pF}[b][b]{\small $\begin{array}{c} (1-p_D)p_F, \\ c_A+c_D \end{array}$}
\psfrag{qF}[tl][tl]{\small $\!\! \begin{array}{c} (1-p_D)(1-p_F),
\\ c_A+c_D \end{array}$}
\psfrag{1}[c][l]{\small $1,c_F+c_D$}
\includegraphics{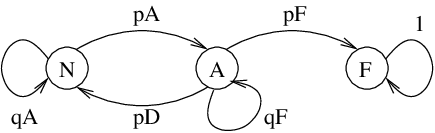} \\[0.1in]
{\small (b) Under a ``Defend'' Action (Control $D$)} \\[0.2in]
\psfrag{qA}[c][c]{\small $1,c_R$}
\psfrag{1}[c][c]{\small $1,c_R$}
\psfrag{pR}[c][c]{\small $p_R,c_R$}
\psfrag{qR}[c][c]{\small $1-p_R,c_F+c_R$}
\includegraphics{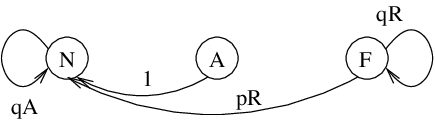} \\
{\small (c) Under a ``Reset'' Action (Control $R$)}\\
\caption{ \label{fig:threeStateModel} A three-state Markov model (a)
  under a ``wait'' action, in which case a system operating normally
  (i.e., in state $N$) will eventually fall under attack and
  subsequently experience a security failure (i.e., transition to
  state $A$ and then to state $F$); (b) under a ``defend'' action,
  which possibly prevents an ongoing attack from causing a security
  failure (i.e., possibly transitions from state $A$ to state $N$);
  and (c) under a ``reset'' action, which can reinstate the system to
  normal operation (i.e., transition to state $N$) no matter the
  current state.}
\end{figure}

We categorize the security environment into only three states, labeled
$N$, $A$ and $F$ to denote ``operating normally,'' ``under attack''
and "security failure,'' respectively; there are similarly only three
candidate controls, labeled $W$, $D$ and $R$ to denote a ``wait''
action, a ``defend'' action, and a ``reset'' action, respectively.
The three diagrams in \FigRef{threeStateModel} summarize the
single-stage transition probabilities/costs under each control action.
For example, under the ``wait'' action (control $W$), self-transitions
in state $N$ represent secure, normal operation of the information
system; however, a transition from state $N$ to $A$ occurs with
per-stage probability $p_A$, representing the beginning of an
intrusion attempt. Thereafter, self-transitions in state $A$ represent
a secure system being under attack; eventually, a transition from
state $A$ to $F$ occurs with per-stage probability $p_F$, representing
the beginning of a security failure that persists indefinitely. Stages
that begin in state $N$ are cost-free, whereas stages that begin in
state $A$ incur a cost of $c_A$ and self-transitions in state $F$
incur a cost of $c_F$. These same probability/cost parameters apply
under the ``defend'' action (control $D$), but with two differences:
firstly, the possible transition from state $A$ back to $N$ (occurring
with per-stage probability $p_D$) represents the successful disruption
of the intrusion attempt; secondly, a cost of $c_D$ is incurred in
addition to the transition cost under control $W$. Finally, under the
``reset'' action (control $R$), the system always returns to normal
operation unless already in the failure state, in which case the
return to normal operation occurs with probability $p_R$; a cost of
$c_R$ is incurred in addition to the transition cost under control
$W$.

Summarizing, our simplest MDP model is described by four probability
parameters and four cost parameters. All probability parameters may
take values strictly between zero and unity; in particular, no value
of one probability restricts the value of another. This reflects the
fact that each probability parameter abstracts a different
characteristic of the intrusion tolerance problem.  More specifically,
probability $p_A$ relates to the frequency with which intrusion
attempts are initiated, where larger values pertain to more aggressive
attackers.  Probability $p_F$ relates to the rate at which an
initiated intrusion will penetrate the security system, where larger
values pertain to more vulnerable systems (or perhaps just
shorter-duration exploits). Probability $p_D$ relates to the success
rate of a well-timed defensive countermeasure in preventing an
otherwise inevitable security failure, where larger values pertain to
more effective response devices. Probability $p_R$ relates to the
success rate of the reset action when reacting to a security failure,
where larger values pertain to a more reliable recovery procedure.
Our analysis will at times also be interested in the special cases
where (i)~probability $p_D$ is exactly zero, corresponding to a
completely useless defend action, and (ii)~probability $p_R$ is
exactly unity, corresponding to a fully reliable reset action.

The four cost parameters may take any set of non-negative values that
satisfy the following inequalities:
\begin{itemize}
\item $c_A < c_F$, or one stage of a security failure incurs more cost
  (e.g., risk of exfiltration) than one stage of an intrusion attempt;
\item $c_D < c_R$, or a reset action incurs more cost (e.g., loss of
  availability) than a defend action; and
\item $c_R \leq c_F$, or one stage of a security failure is at least
  as costly as invoking the reset action.
\end{itemize}
Note that, because costs $c_A$ or $c_D$ are always non-negative, these
conditions imply that costs $c_F$ and $c_R$ are always strictly
positive.

Subject to design in a MDP model is the policy by which, in each
stage, all available information maps into each selected control,
typically seeking to minimize the total cost accrued over a large
number of stages. Under such a penalty function, the above
restrictions on cost parameters are quite natural, reflecting an
intuition that it be preferable to take actions $W$, $D$ or $R$ when
in states $N$, $A$ or $F$, respectively. We will see (in
\SecRef{PerDet}) that this natural policy is not necessarily the best
choice---the uncertainty (expressed via the probability parameters)
must also be accounted for. Furthermore, upon moving to the realism of
imperfect detectors (in \SecRef{ModExt}), we may never be certain
about the true system state and must instead drive the decision
process with a (sometimes incorrect) state estimate. Regardless of
detector assumptions, design of the control policy must manage
inherent tradeoffs between total overhead due to defensive
countermeasures or system resets (with per-stage costs $c_D$ and
$c_R$, respectively) and total risks due to ongoing intrusions or
security failures (with per-stage costs $c_A$ or $c_F$, respectively);
moreover, polices that fail to account for all of the underlying
uncertainty, in both state evolution and state estimation, are also
unlikely to achieve the best long-term balance between these two types
of penalties.


\section{Analysis: Perfect Detectors \label{sec:PerDet}}
We now proceed to analyze the main MDP model described in
\SecRef{MainModel} under the assumption of perfect detectors, in which
case the optimal policy admits an analytical solution. It is shown to
always be one of only three alternative policies, and which of the
three is characterized as the solution to a particular system of
inequalities. The dependence on model parameters is complicated, so we
develop an understanding of these inequalities in a piece-meal
fashion.  In particular, we first examine a series of special cases of
the main model (e.g., fully reliable reset actions, cost-free
intrusion attempts), reaching the most general understanding in the
final subsection.  Removing the unrealistic assumption of perfect
detectors requires the extended modeling formalism of a
partially-observable Markov decision process (POMDP), which is
discussed (among other extensions) in \SecRef{ModExt}.

\subsection{Characterization of the Optimal Policy}
Mathematically, perfect detection means that, in each stage $k = 0, 1,
\ldots$, the current state $x_k \in \{ N,A,F\}$ is revealed before the
control decision $u_k\in\{ W,D,R\}$ must be made. Of course, the next
state $x_{k+1}$ remains uncertain until action $u_k$ is applied and
the process proceeds to the subsequent decision stage.  Let
$g(x_k,u_k,x_{k+1})$ denote the cost associated with the outcome of
stage $k$ e.g., in our model, it may be any one of the different
transition costs provided in \FigRef{threeStateModel}. We desire a
control policy that minimizes the \emph{infinite-horizon
  average-cost-per-stage}
\begin{equation}
\label{equ:optCrit}
\lambda = \lim_{T\rightarrow \infty}\left( \frac{1}{T}\sum_{k=0}^T
  g(x_k,u_k,x_{k+1}) \right),
\end{equation}
which under certain technical conditions (satisfied by our model) is
known to be achievable with a \emph{stationary}, or stage-invariant,
policy \cite{Ber95:DPOC1}.  This means, in our model, that the optimal
policy is one of the twenty-seven functions of the form $\mu:\{N,A,F\}
\rightarrow \{W,D,R\}$, each a different mapping from the state space
to the control space by which every single-stage decision $u_k =
\mu(x_k)$ is to be made.

Given a particular policy $\mu$, equations are provided in
\cite{Ber95:DPOC1} (c.f., Proposition~4.1(c) in Chapter~7) to derive the
associated average-cost-per-stage $\lambda^\mu$ in terms of the given
model parameters. However, under the assumptions on cost parameters
stated in \SecRef{MainModel}, it is easily seen that most of the
twenty-seven candidate policies may be quickly dismissed.  Firstly,
because the only hope of exiting the failure state is by invoking the
reset action (and $c_R \leq c_F$), we need only consider policies that
map state $F$ to action $R$. Secondly, because the defend action has
no bearing on transitions out of normal operation (and $c_D \geq 0$),
we need \emph{not} consider policies that map state $N$ to action $D$.
This leaves only six candidate policies, three mapping state $N$ to
the wait action $W$ and three mapping state $N$ to the reset action
$R$---let us first evaluate the former three policies, mapping state
$A$ to different actions.
\begin{enumerate}
\item \emph{Wait-upon-attack} $\mu^W$, or taking action $W$ when in
  state $A$ and incurring cost $c_A$ per stage until the transition to
  state $F$ occurs, achieving
\begin{equation}
  \label{equ:lambdaW}
  \lambda^W = \frac{p_A\left[p_Rc_A + p_F(1-p_R)c_F+ p_Fc_R\right]}
{p_Ap_F + p_R(p_A+p_F)}.
\end{equation}
\item \emph{Defend-upon-attack} $\mu^D$, or taking action $D$ when in
  state $A$ and incurring cost $c_A + c_D$ per stage until the
  transition to state $N$ or to state $F$ occurs, achieving
\begin{equation}
\label{equ:lambdaD}
\lambda^D = \frac{G^W(1-p_D) + p_Ap_Rc_D}{Q^W + p_D
\left[ p_R(1-p_F)-p_Fp_A\right]}
\end{equation}
with $G^W$ and $Q^W$ denoting the numerator and denominator,
respectively, of $\lambda^W$ in \equRef{lambdaW}.
\item \emph{Reset-upon-attack} $\mu^R$, or taking action $R$
  immediately upon exiting state $N$, achieving
\begin{equation}
\label{equ:lambdaR}
\lambda^R = \frac{p_Ac_R}{1+p_A}.
\end{equation}
\end{enumerate}
The remaining policies, namely those mapping state $N$ to action $R$,
each achieve an average-cost-per-stage of $c_R$ and can thus be
dismissed by virtue of \equRef{lambdaR}.

In summary, the optimal policy for the MDP model of \SecRef{MainModel}
is one of only three candidate policies, each mapping the normal state
$N$ to the wait action $W$ and the failure state $F$ to the reset
action $R$ but differing in the action taken when in the attack state
$A$. Observe that policies $\mu^W$ or $\mu^D$, employing the reset
action only after the system enters the failure state $F$, represent
favoring a reactive recovery strategy whereas policy $\mu^R$,
employing the reset action immediately after the system exits the
normal state $N$, represents favoring a proactive recovery strategy.
The average-cost-per-stage performance of these three policies are
given by \equRefs{lambdaW}{lambdaR}, each a closed-from expression in
terms of the model parameters. Thus, our characterization of the
optimal policy boils down to making just three direct comparisons
i.e., the ``wait-under-attack'' policy $\mu^W$ is optimal if and only
if $\lambda^W < \lambda^R$ and $\lambda^W < \lambda^D$; otherwise, the
``defend-under-attack'' policy $\mu^D$ is optimal if and only if
$\lambda^D < \lambda^R$; otherwise, the ``reset-under-attack'' policy
$\mu^R$ is optimal. The following inequalities express these
comparisons in terms of the model parameters.

\noindent
$\bullet$ Comparison $\lambda^W < \lambda^R$ holds if and only if
\begin{equation}
\label{equ:CompWtoR}
\begin{array}{l}
(1+p_A)\left[ p_Rc_A + p_F(1-p_R)c_F\right] < \\[0.05in]
\qquad \qquad \qquad \quad \qquad \qquad \qquad
\left[ p_R(p_A + p_F)-p_F\right] c_R.
\end{array}
\end{equation}

\noindent
$\bullet$  Comparison $\lambda^W < \lambda^D$ holds if and only if
\begin{equation}
\label{equ:CompWtoD}
\begin{array}{l}
(1+p_A)p_D\left[ p_Rc_A + p_F(1-p_R)c_F+p_Fc_R\right] < \\[0.05in]
\qquad \qquad \qquad \qquad \qquad \quad \:\:\:
\left[ p_Ap_F + p_R(p_A+p_F)\right] c_D.
\end{array}
\end{equation}

\noindent
$\bullet$ Comparison $\lambda^D < \lambda^R$ holds if and only if
\begin{equation}
\label{equ:CompDtoR}
\begin{array}{l}
(1+p_A)\left[ (1-p_D)\left[ p_Rc_A + p_F(1-p_R)c_F\right] +
p_Rc_D\right] < \\[0.05in]
\qquad \qquad \quad \:\:\: \left[ p_R(p_A+p_D) - p_F(1-p_D)(1-p_R)
\right] c_R.
\end{array}
\end{equation}

\subsection{Some First Insights}
Notice that comparing policies \equRefs{lambdaW}{lambdaR} in terms of
the Mean-Time-To-Failure (MTTF) or Mean-Time-To-Repair (MTTR) would
not fully capture the long-term cost tradeoffs. This is easiest to see
in the special case of a fully reliable reset action, or when
probability $p_R$ is unity. In this case, a failure (followed by an
immediate reset) occurs on-average once every $1 + 1/p_A + 1/p_F$
stages under policy $\mu^W$ and once every $1 + 1/p_A + (p_A +
p_D)/[p_Ap_F(1-p_D)]$ stages under policy $\mu^D$.  Moreover, under
policy $\mu^R$, the MTTF is technically infinite but, proactively, the
reset action is invoked on-average once every $1 + 1/p_A$ stages.  So,
regardless of the values of the remaining three probability
parameters, policy $\mu^D$ results in the longest MTTF. (Here, for ease of
exposition, each discrete stage is assumed to consume the same
duration of time; extension to non-uniform time discretization will be
discussed in \SecRef{ModExt}.) In contrast, it is readily seen that the ordering
of \equRefs{lambdaW}{lambdaR} also depends on the values for cost parameters
$c_A$, $c_D$ and $c_R$ (as well as cost parameter $c_F$ if the reset action
is unreliable, or if probability $p_R$ is strictly less than unity).

Next, consider the limiting case in which the per-stage failure cost
$c_F$ approaches infinity e.g., a security failure translates into
loss-of-reputation or loss-of-life and is, in turn, orders of
magnitude more costly than any other event. Then, any policy with
which there is even the smallest yet non-zero chance of incurring cost
$c_F$ is infinitely penalized.  \EquRefs{lambdaW}{lambdaR} show that
$\mu^R$ is the only policy with zero such chance \emph{unless}
probability $p_R$ is unity (i.e., unless the reset action is fully
reliable), in which case all three policies remain on the table.
Thus, we have just deduced the obvious conclusion that a proactive
recovery strategy (represented in our model by policy $\mu^R$) is best
if (i)~a security failure is equated with catastrophic consequences
and (ii)~reset actions do not always recover from a security
failure. However, despite the infinite failure cost, a proactive
recovery strategy is \emph{not} necessarily preferable if the reset
action is fully reliable.  Now, these conclusions are also premised on
the idealization that we can perfectly (and immediately) detect the
onset and completion of each intrusion attempt; as will be discussed
in \SecRef{ModExt}, the choice between reactive versus proactive
recovery gets clouded by the reality of imperfect detectors.

The comparisons via \equRefs{lambdaW}{lambdaR} capture subtleties
beyond just explicit dependence on the cost parameters. Consider, for
example, the special case in which probability parameter $p_D$ is
near-unity and cost parameter $c_D$ is exactly zero (i.e., the
defensive action is both highly effective and incurs no costly
overhead). Is policy $\mu^D$ then necessarily optimal?  Notice that
parameters $p_D$ and $c_D$ appear only in the expression for
$\lambda^D$ in \equRef{lambdaD}. While the numerator certainly
decreases as probability $p_D$ increases, the same cannot be said for
penalty $\lambda^D$ on the whole. In particular, if probability $p_R$
is less than $p_Ap_F/(1-p_F)$, the denominator of $\lambda^D$ also
decreases as $p_D$ increases. Thus, it need not be the case that a
cost-free, highly-effective defensive action should always be
employed: the optimality of policy $\mu^D$ may somehow also depend on
the reliability of the reset action in comparison to attack dynamics.

\begin{figure}[!t]
\centering
\psfrag{0}[c][c]{\small $0$}
\psfrag{1}[c][c]{\small $1$}
\psfrag{pR}[c][c]{reset reliability (probability $p_R$)}
\psfrag{p1}[r][r]{$\frac{p_F(1-p_D)}{p_F(1-p_D) + p_A + p_D}$}
\psfrag{p2}[r][r]{$\frac{p_F}{p_F + p_A}$}
\psfrag{T1}[l][l]{\small $\begin{array}{c} \mbox{\sc Insufficiency}
\\[0.05in] \mbox{(policy $\mu^R$ is optimal)} \end{array}$}
\psfrag{T2}[l][l]{\small $\begin{array}{c} \mbox{\sc Weak Sufficiency}
\\[0.05in] \mbox{(either policy $\mu^D$} \\ \mbox{or $\mu^R$ is optimal)}
\end{array}$}
\psfrag{T3}[l][l]{\small $\begin{array}{c} \mbox{\sc Strong Sufficiency}
\\[0.05in] \mbox{(either policy $\mu^W$, $\mu^D$} \\ \mbox{or $\mu^R$ is
optimal)} \end{array}$}
\includegraphics{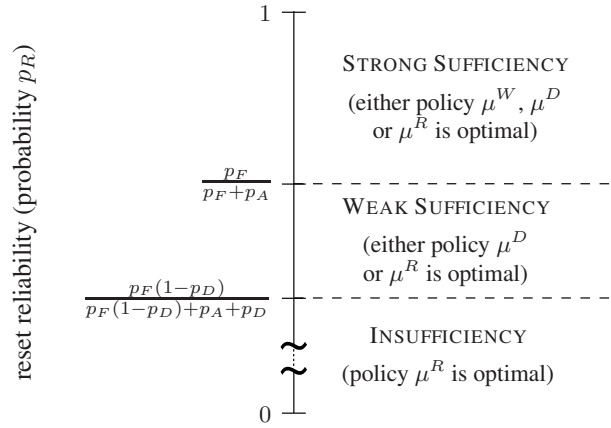}
\caption{Notions of ``sufficiently reliable'' reset actions.
\label{fig:resetSuff}}
\end{figure}

Indeed, the ordering of \equRefs{lambdaW}{lambdaR} generally depends
on all model parameters, the probabilities as well as the costs. This
general dependence is captured most succinctly by
\equRefs{CompWtoR}{CompDtoR}, or the system of inequalities expressed
in terms of the model parameters. These will be examined from
different viewpoints in the following subsections. For now, we will
introduce the notion of ``sufficiently reliable'' reset actions by
focusing on the inequalities \equRef{CompWtoR} and \equRef{CompDtoR}
that involve the policy $\mu^R$.  Notice that the right-hand-sides of
these two inequalities are the only two terms in all of
\equRefs{CompWtoR}{CompDtoR} that may take negative values.  Because
the associated left-hand-sides cannot be negative, the comparisons
become moot when these terms are in fact negative. Specifically,
regardless of the values taken by the four cost parameters, it follows
from \equRef{CompWtoR} that
\begin{equation}
\label{equ:strongNotion}
\mbox{if } p_R < \frac{p_F}{p_F + p_A}, \quad \mbox{then }
\lambda^R < \lambda^W
\end{equation}
and from \equRef{CompDtoR} that
\begin{equation}
\label{equ:weakNotion}
\mbox{if } p_R < \frac{p_F(1-p_D)}{p_F(1-p_D) + p_A + p_D}, \quad
\mbox{then } \lambda^R < \lambda^D.
\end{equation}
Also recognizing that
the right-hand-side of \equRef{strongNotion} is greater than or equal
to the right-hand-side of \equRef{weakNotion} for any value of
probability $p_D$, we can identify two distinct notions of
``sufficiently reliable'' reset actions.  As \FigRef{resetSuff}
illustrates, the strong notion is when probability $p_R$ exceeds the
threshold in \equRef{strongNotion}, in which case all three policies
remain on the table until the cost parameters enter the picture;
otherwise, the weak notion is when probability $p_R$ exceeds the
threshold in \equRef{weakNotion}, in which case only policies $\mu^D$
and $\mu^R$ remain; and otherwise only policy $\mu^R$ remains and is
thus optimal, rendering the cost parameters entirely irrelevant.

It is not surprising that the reliability of the reset action is
another key factor in deciding between a reactive or proactive
recovery strategy. Far from obvious, however, is the manner in which
the other probability parameters influence the decision through their
impact on the notions of ``sufficiently reliable'' reset actions. In
particular, we see that sufficiency is more difficult to obtain, in
the sense of requiring reset actions to have a higher probability
$p_R$, in scenarios with probability $p_A$ much smaller than
probability $p_F$ i.e., infrequent and/or fast attacks. Moreover, the
region of weak sufficiency diminishes with decreasing probability
$p_D$ i.e., less effective defend actions; in the limit of a useless
defend action (when $p_D$ is zero), not only does the region of weak
sufficiency disappear entirely but we also see from inequality
\equRef{CompWtoD} that policy $\mu^D$ can be dismissed even under
strong sufficiency. Altogether, when facing infrequent and/or fast
attacks (and especially when armed only with defend actions of lesser
effectiveness), the region of insufficiency is likely to dominate and,
in turn, proactive recovery is likely the better strategy.

\subsection{The Special Case of Reliable Reset  ($p_R = 1$)}
This subsection will examine the inequalities
\equRefs{CompWtoR}{CompDtoR} in the special case of a fully reliable
reset action, or when probability $p_R$ is exactly unity. As discussed
in the preceding subsection, this guarantees a policy selection
problem that is both in the region of strong sufficiency (see
\FigRef{resetSuff}) and independent of cost parameter $c_F$; however,
values for the three other cost parameters do enter into the
considerations.

Consider first the comparison between $\lambda^W$ and $\lambda^R$ in
\equRef{CompWtoR}, which in this subsection specializes to the
inequality $(1+p_A)c_A < p_Ac_R$. That is, $\lambda^W$ is less than
$\lambda^R$ if and only if the cost ratio $c_A/c_R$ is less than the
fraction $p_A/(1+p_A)$. This is arguably surprising: despite its
appearance in \equRef{lambdaW} (even with $p_R = 1$), probability
$p_F$ turns out to have no bearing on the comparison between policy
$\mu^W$ and policy $\mu^R$.  In particular, the ordering of
$\lambda^W$ and $\lambda^R$ is unchanged even as $p_F$ is near-unity,
or the case where each intrusion attempt almost immediately results in
a security failure (e.g., a flash worm \cite{SMP04:Flash}). An equivalent
interpretation, but from the perspective of holding cost parameters
fixed, is that policy $\mu^W$ is preferable to policy $\mu^R$ only in
the case of sufficiently frequent attacks (i.e., only for large enough
$p_A$ such that $p_A/(1+p_A)$ exceeds the ratio $c_A/c_R$)---in the
case of less frequent attacks or simply when $c_A/c_R$ exceeds $1/2$,
policy $\mu^R$ is preferable.

The preceding comparison between $\lambda^W$ and $\lambda^R$ can be
expressed concisely by the inequalities
\begin{equation}
\label{equ:ruleWvsR}
\lambda^R
\begin{array}{c}
\mbox{\small choose $\mu^W$} \\[-0.1cm]
> \\
< \\[-0.1cm]
\mbox{\small choose $\mu^R$}
\end{array}
\lambda^W \quad \Longleftrightarrow \quad
\frac{p_A}{1+p_A}
\begin{array}{c}
\mbox{\small choose $\mu^W$} \\[-0.1cm]
> \\
< \\[-0.1cm]
\mbox{\small choose $\mu^R$}
\end{array}
\frac{c_A}{c_R} \quad .
\end{equation}
Similarly, the comparison between $\lambda^W$ and $\lambda^D$
in
\equRef{CompWtoD} specializes to the inequalities
\begin{equation}
\label{equ:ruleWvsD}
\frac{c_D}{c_R}
\begin{array}{c}
\mbox{\small choose $\mu^W$} \\[-0.1cm]
> \\
< \\[-0.1cm]
\mbox{\small choose $\mu^D$}
\end{array}
\left( \frac{p_D(1+p_A)}{p_A+p_F(1+p_A)} \right) \frac{c_A}{c_R} + y_0
\end{equation}
with
$$
y_0 = \frac{(1+p_A)p_Fp_D}{p_A+p_F(1+p_A)}.
$$
Finally, the comparison 
in
\equRef{CompDtoR} specializes to
\begin{equation}
\label{equ:ruleDvsR}
\frac{c_D}{c_R}
\begin{array}{c}
\mbox{\small choose $\mu^R$} \\[-0.1cm]
> \\
< \\[-0.1cm]
\mbox{\small choose $\mu^D$}
\end{array}
-(1-p_D)\frac{c_A}{c_R} + \frac{p_A+p_D}{1+p_A}
\quad .
\end{equation}

Observe that probability $p_A$ affects all three pairwise comparisons,
probability $p_D$ affects only comparisons \equRef{ruleWvsD} and
\equRef{ruleDvsR}, while probability $p_F$ affects only comparison
\equRef{ruleWvsD}.  Viewing the (normalized) costs $c_A/c_R$ and
$c_D/c_R$ as degrees-of-freedom, \equRefs{ruleWvsR}{ruleDvsR} form a
system of linear inequalities with coefficients depending on the
values of probabilities $p_A$, $p_F$ and $p_D$.  In particular, these
normalized costs comprise a two-dimensional design space,
and each inequality alone defines a partition of this space in
accordance with which of the two associated policies is preferable. In
turn, all inequalities together define the partition in accordance
with which of the three policies is optimal.

\begin{figure}[!t]
\centering
\psfrag{title}{}
\psfrag{xlabel}[t][b]{\large $c_A/c_R$}
\psfrag{ylabel}[b][t]{\large $\begin{array}{c}
c_D/c_R \\[-0.05in] \textcolor{white}{A} \end{array}$}
\psfrag{y0}[t][b]{$0$}
\psfrag{1}{$1$}
\psfrag{x1}[t][t]{\large $\frac{p_A}{1+p_A}$}
\psfrag{x2}[t][t]{\large $\frac{p_A+p_D}{(1-p_D)(1+p_A)}$}
\psfrag{y1}[r][r]{$y_0$}
\psfrag{y2}[r][r]{$p_D$}
\psfrag{y3}[r][r]{\large $\frac{p_A+p_D}{1+p_A}\!$}
\psfrag{s}[b][b]{}
\psfrag{choosew}[b][b]{$\!\!\!\!$ choose $\mu^W$}
\psfrag{choosed}{$\!\!\!\!$ choose $\mu^D$}
\psfrag{chooser}{choose $\mu^R$}
\resizebox{1.5in}{!}{\includegraphics*[140,390][290,540]{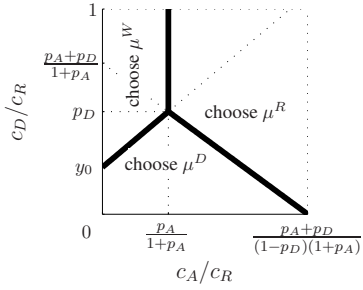}}
\caption{\label{fig:relResetPartitions} The partitions characterizing the
  optimal policy for the MDP model in \FigRef{threeStateModel} in the
  special case of a fully reliable reset action ($p_R = 1$).}
\end{figure}

\FigRef{relResetPartitions} summarizes the optimal partition
graphically, where we define $y_0$ as in \equRef{ruleWvsD}.
Consistent with intuition, the optimal policy is (i)~to
reset-under-attack $\mu^R$ when both the attack cost $c_A$ and
the defend cost $c_D$ are large, (ii)~to defend-under-attack $\mu^D$ when
both $c_A$ and $c_D$ are small and (iii)~to wait-under-attack $\mu^W$ when
$c_A$ is small but $c_D$ is large. Less intuitive is what the optimal
policy is when the costs take mid-range values and how the probability
parameters influence the matter. Notice first that the
three regions always meet at the point where $c_A/c_R = p_A/(1+p_A)$
and $c_D/c_R = p_D$. It's also seen that policy $\mu^W$ cannot be
optimal if $c_A/c_R$ exceeds $p_A/(1+p_A)$ or $c_D/c_R$ is below
$y_0$, while policy $\mu^D$ cannot be optimal if $c_A/c_R$ exceeds
$(p_A + p_D)/[(1-p_D)(1+p_A)]$ or $c_D/c_R$ exceeds $p_D$.  As
probability $p_F$ increases (i.e., as attacks more quickly bring about
failure), the boundary between policies $\mu^W$ and $\mu^D$ pivots
clockwise about the central meeting point.  Finally, the sensitivity
of the optimal partition to such changes in probabilities $p_D$ or
$p_F$ is also a function of probability $p_A$ (i.e., of the frequency
with which attacks occur).

It is also instructive to consider the limiting cases of probability
$p_D$ approaching zero and unity, which correspond to the useless and
the perfect defend action, respectively. Notice that the region for
policy $\mu^D$ reduces (expands) in size as $p_D$ decreases
(increases), vanishing entirely as $p_D$ approaches zero (replacing
entirely the region for policy $\mu^R$ as $p_D$ approaches
unity)---this is not surprising, considering that a useless (perfect)
defend action is essentially equivalent to an expensive wait (a cheap
preemptive reset) action. However, even as $p_D$ approaches unity, the
region of policy $\mu^W$ does \emph{not} entirely vanish because the
point $y_0$ approaches the value $p_F(1+p_A)/(p_F(1+p_A) + p_A)$,
which is strictly less than unity by an amount that grows with
increasing probability $p_A$.  Thus, even when the defend action is
extremely effective and especially against higher-frequency attacks,
it is possible (e.g., when $c_D/c_R$ exceeds $y_0$ and $c_A/c_R$ is
zero) that policy $\mu^W$ remains preferable to policy $\mu^D$ i.e.,
that a reactive recovery strategy without the use of the near-perfect
defend action is best.

\begin{figure}[!h]
\centering
\psfrag{xlabel}[t][b]{\Large $c_A/c_R$}
\psfrag{ylabel}[m][b]{\Large $c_D/c_R$}
\psfrag{0}[c][c]{$0$}
\psfrag{1}[c][c]{$1$}
\psfrag{x1}[c][t]{\large $\:\:\:\:\:\frac{p_A}{1+p_A}$}
\psfrag{x2}[c][c]{}
\psfrag{y0}[l][r]{$\: 0$}
\psfrag{y1}[c][c]{}
\psfrag{y2}[c][tr]{$p_D$}
\psfrag{y3}[c][c]{}
\psfrag{z1}[c][b]{\large $\frac{p_A}{1+p_A}\:\:$}
\vspace{0.08in}
\begin{tabular}{ccc}
\hspace*{-0.1in}
\psfrag{title}[c][c]{\Large Cost-per-Stage}
\psfrag{zlabel}[b][t]{\Large $\lambda^*/c_R$}
\resizebox{1in}{!}{\includegraphics*[115,440][250,545]{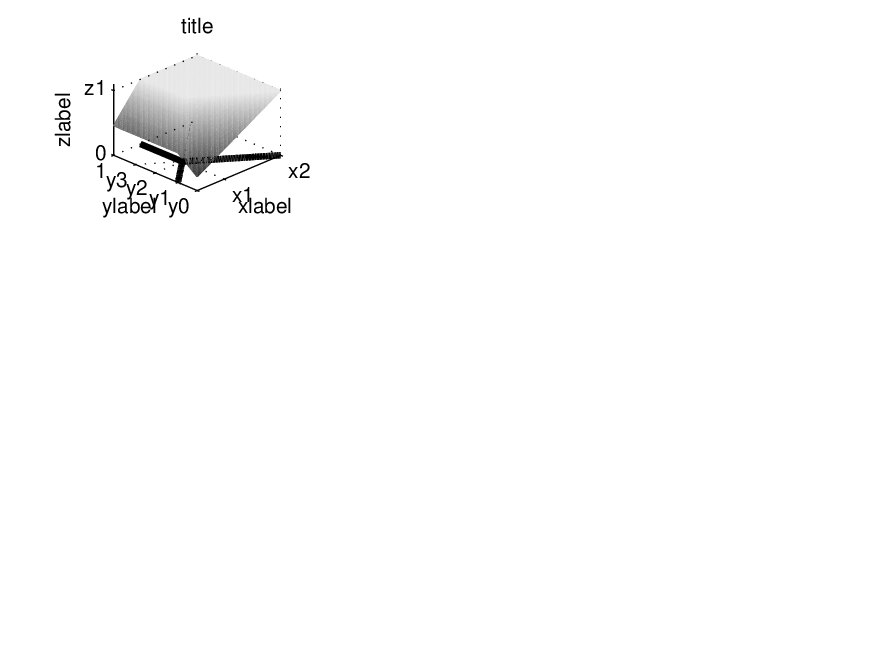}}
&
\hspace*{-0.1in}
\psfrag{title}[c][c]{\Large Maintenance Overhead}
\psfrag{zlabel}[b][t]{\Large $\lambda^*_M/c_R$}
\resizebox{1in}{!}{\includegraphics*[115,440][250,545]{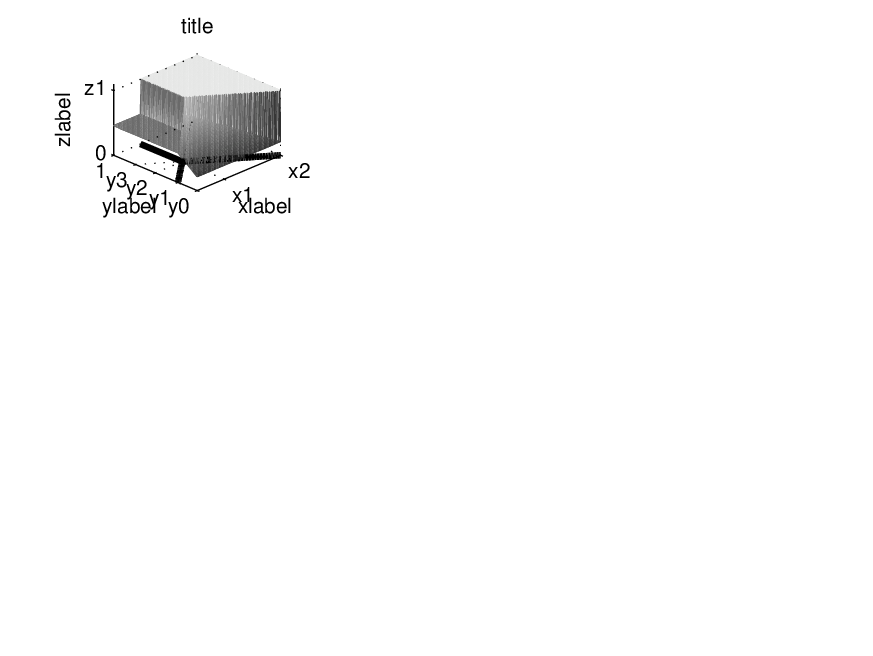}}
&
\hspace*{-0.1in}
\psfrag{title}[c][c]{\Large Failure Risk}
\psfrag{zlabel}[b][t]{\Large $\lambda^*_F/c_R$}
\resizebox{1in}{!}{\includegraphics*[115,440][250,545]{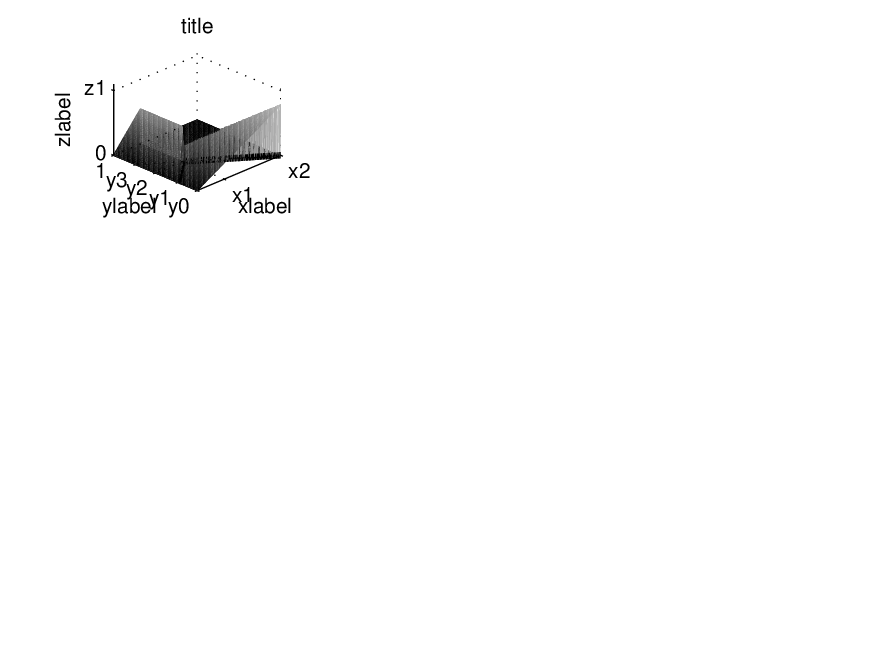}} \\
\hspace*{-0.1in}
{\small (a)} &
\hspace*{-0.1in}
{\small (b)} &
\hspace*{-0.1in}
{\small (c)} \\
\end{tabular}
\caption{\label{fig:relResetPerformance} The (a) performance of the
  optimal policy in \FigRef{relResetPartitions} and its tradeoff
  between (b)~maintenance overhead and (c)~failure risk.}
\end{figure}

\FigRef{relResetPerformance}(a) plots the achieved
average-cost-per-stage $\lambda^*$ (normalized by the reset cost
$c_R$). It is seen to increase linearly as either $c_A$ or $c_D$
increases \emph{unless} these costs are such that policy $\mu^R$ is
optimal, in which case $\lambda^*$ plateaus at $\lambda^R$.
\FigRef{relResetPerformance}(b) and \FigRef{relResetPerformance}(c)
illustrate that $\lambda^*$ is, in general, the sum of two types of
security-related penalties. Here, we have denoted the maintenance
overhead by $\lambda_M^*$, which is the sum of terms involving costs
$c_D$ or $c_R$, and the failure risk by $\lambda_F^*$, which is the
sum of terms involving costs $c_A$ or $c_F$. For example, $\lambda^W$
decomposes into the sum of
$$
\lambda^W_M = \frac{p_Ap_Fc_R}{p_A + p_F(1+p_A)} \mbox{ and }
\lambda^W_F = \frac{p_Ac_A}{p_A + p_F(1+p_A)},
$$
whereas $\lambda^R$ decomposes into $\lambda^R_M = \lambda^R$ and
$\lambda^R_F = 0$.

\subsection{The Special Case of Cost-Free Attack ($c_A = 0$)}
This subsection will examine the inequalities
\equRefs{CompWtoR}{CompDtoR} in the special case of cost-free attacks,
or when the per-stage cost $c_A$ associated with intrusion attempts is
exactly zero. Recall that, in our model, the per-stage failure cost
$c_F$ associated with allowing an intrusion to complete is always
nonzero and greater than the two control-related costs, namely $c_D$
and $c_R$. As in the preceding subsection, we begin by specializing
the inequalities \equRefs{CompWtoR}{CompDtoR} into a more concise form
in terms of two normalized costs. However, now that probability $p_R$
need not be unity and thus cost $c_F$ is pertinent, the normalized
costs are $c_D/c_F$ and $c_R/c_F$, each quantifying the expense of the
different control actions relative to each stage of security failure.
We will again view these normalized costs as a two-dimensional design
space and express the optimal policy as a specific partition that
depends on the probability parameters $p_A$, $p_F$, $p_D$ and
$p_R$. What will also emerge, however, are refined notions of
''sufficiently reliable'' reset actions (in comparison to
\FigRef{resetSuff}). In particular, recognizing that $c_R$ is never
greater than $c_F$, we obtain tighter conditions on the probability
parameters for which policies $\mu^W$ or $\mu^D$ can be dismissed
without consideration of the costs.

We begin by substituting $c_A = 0$ into the inequalities
\equRefs{CompWtoR}{CompDtoR} and rearranging terms. Under the
assumption of strong efficiency, the comparison between $\lambda^W$
and $\lambda^R$ specializes to the inequalities
\begin{equation}
\label{equ:ruleWvsR2}
\frac{c_R}{c_F}
\begin{array}{c}
\mbox{\small choose $\mu^W$} \\[-0.1cm]
> \\
< \\[-0.1cm]
\mbox{\small choose $\mu^R$}
\end{array}
\frac{(1+p_A)p_F(1-p_R)}{p_R(p_A+p_F) - p_F} \quad .
\end{equation}
Similarly, the comparison between $\lambda^W$ and $\lambda^D$ in
\equRef{CompWtoD} specializes to the inequalities
\begin{equation}
\label{equ:ruleWvsD2}
\frac{c_D}{c_F}
\begin{array}{c}
\mbox{\small choose $\mu^W$} \\[-0.1cm]
> \\
< \\[-0.1cm]
\mbox{\small choose $\mu^D$}
\end{array}
\left( \frac{p_D(1+p_A)p_F}{p_Ap_F+p_R(p_A+p_F)} \right) \frac{c_R}{c_F} + y_1
\end{equation}
with
$$
y_1 = \frac{(1+p_A)p_Fp_D(1-p_R)}{p_Ap_F+p_R(p_A+p_F)}.
$$
Finally, the comparison 
in
\equRef{CompDtoR} specializes to
\begin{equation}
\label{equ:ruleDvsR2}
\frac{c_D}{c_F}
\begin{array}{c}
\mbox{\small choose $\mu^R$} \\[-0.1cm]
> \\
< \\[-0.1cm]
\mbox{\small choose $\mu^D$}
\end{array}
m_1 \frac{c_R}{c_F} - \frac{p_F(1-p_D)(1-p_R)}{p_R}
\end{equation}
with
$$
m_1 =\frac{p_R(p_A+p_D) - p_F(1-p_D)(1-p_R)}{(1+p_A)p_R} .
$$
Recall that, under the assumption of weak sufficiency, policy $\mu^W$
can be immediately dismissed and thus only comparison
\equRef{ruleDvsR2} remains relevant.

\begin{figure}[!b]
\centering
\psfrag{title}{}
\psfrag{xlabel}[c][c]{\large $c_R/c_F$}
\psfrag{ylabel}[b][t]{\large $c_D/c_F$}
\psfrag{0}[t][b]{$0$}
\psfrag{y0}[t][b]{$0$}
\psfrag{1}{$1$}
\psfrag{x1}[c][c]{$\:\:\:\: x_1$}
\psfrag{x2}[c][c]{$\:\: x_2$}
\psfrag{y1}[c][r]{$y_1\:$}
\psfrag{y2}[r][r]{$p_Dx_2$}
\psfrag{y3}[r][rt]{$p_D$}
\psfrag{y4}[r][r]{$y_2$}
\psfrag{choosew}[t][t]{$\!\!\!\!$ choose $\mu^W$}
\psfrag{choosed}{$\!\!\!\!\!$ choose $\mu^D$}
\psfrag{chooser}{$\!\!\!\!\!\!$ choose $\mu^R$}
\begin{tabular}{cc}
\hspace*{0.05in}
\resizebox{1.5in}{!}{\includegraphics*[140,390][290,540]{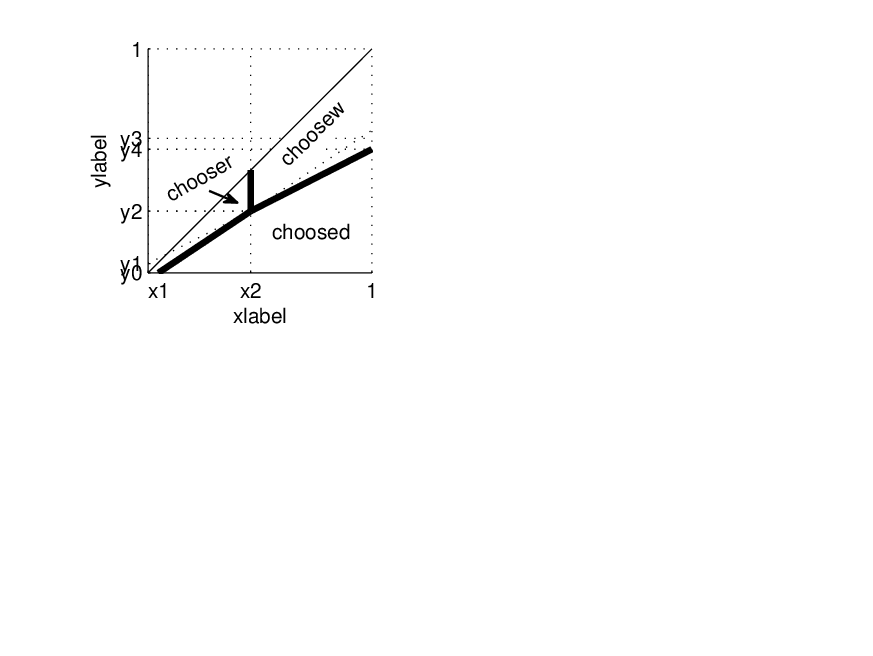}} &
\psfrag{ylabel}[c][t]{\large $c_D/c_F$}
\psfrag{chooser}[c][t]{choose $\mu^R$}
\psfrag{y3}[r][r]{$p_D$}
\psfrag{y5}[c][c]{$y_3$}
\resizebox{1.5in}{!}{\includegraphics*[140,390][290,540]{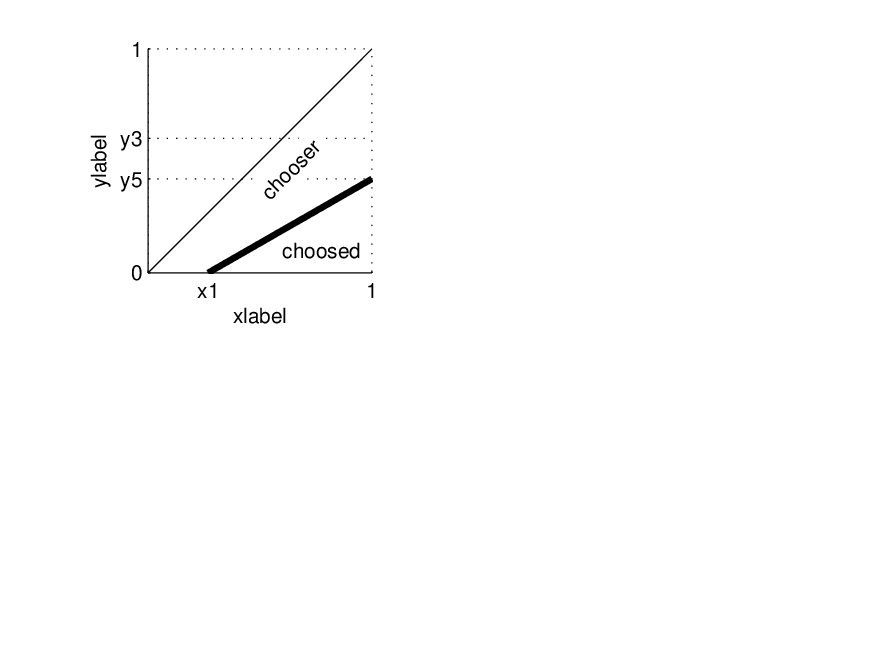}} \\
\hspace*{0.05in}
{\small (a)} & {\small (b)}
\end{tabular}
\caption{\label{fig:freeAttackPartitions} The partitions
  characterizing the optimal policy for the MDP model in
  \FigRef{threeStateModel} in the special case of a cost-free attack
  ($c_A = 0$) and under (a)~strong sufficiency or (b)~weak sufficiency
  (as defined in \FigRef{resetSuff}).}
\end{figure}

\begin{figure}[!h]
\centering
\psfrag{xlabel}[t][b]{\Large $c_R/c_F$}
\psfrag{ylabel}[t][b]{\Large $c_D/c_F$}
\psfrag{0}[c][c]{$0$}
\psfrag{1}[c][c]{$1$}
\psfrag{x1}[c][t]{}
\psfrag{x2}[c][c]{$x_2$}
\psfrag{y0}[l][r]{$\: 0$}
\psfrag{y1}[c][c]{}
\psfrag{y2}[r][tr]{$p_Dx_2$}
\psfrag{y3}[c][c]{}
\psfrag{z1}[c][b]{\large $\frac{p_A}{1+p_A}\:\:$}
\vspace{0.08in}
\begin{tabular}{ccc}
\hspace*{-0.1in}
\psfrag{title}[c][c]{\Large Cost-per-Stage}
\psfrag{zlabel}[b][t]{\Large $\lambda^*/c_R$}
\resizebox{1in}{!}{\includegraphics*[115,440][250,545]{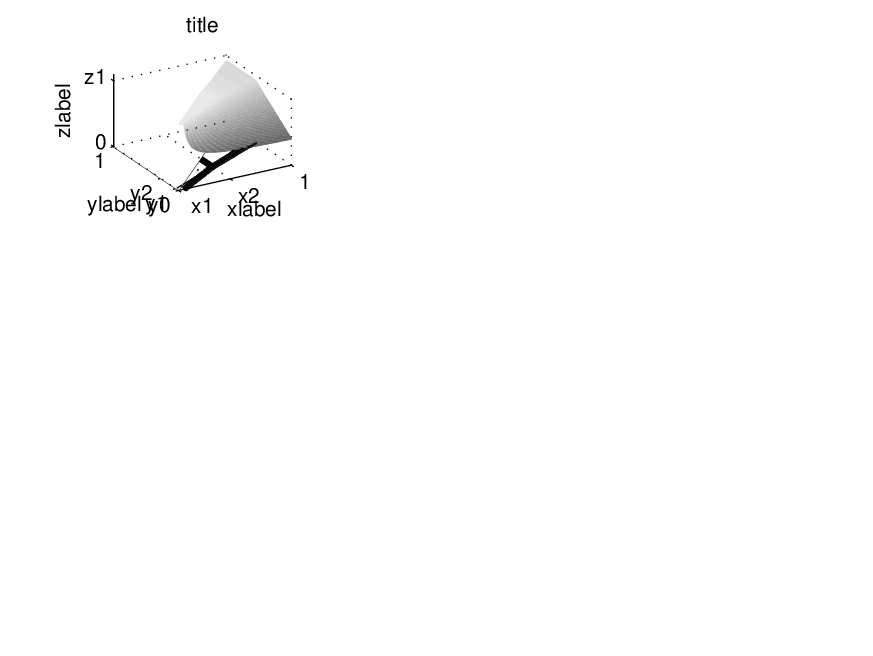}}
&
\hspace*{-0.1in}
\psfrag{title}[c][c]{\Large Maintenance Overhead}
\psfrag{zlabel}[b][t]{\Large $\lambda^*_M/c_R$}
\resizebox{1in}{!}{\includegraphics*[115,440][250,545]{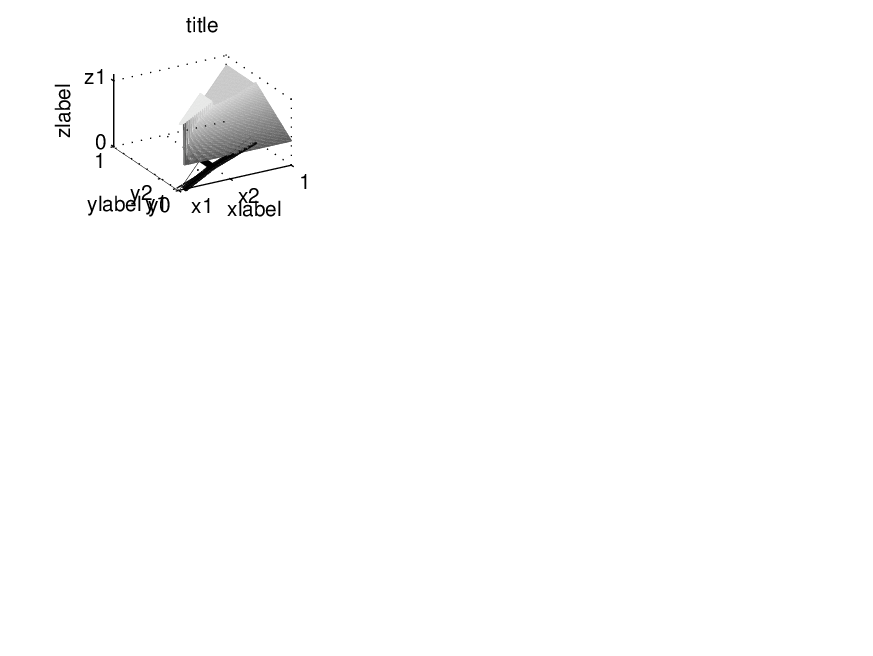}}
&
\hspace*{-0.1in}
\psfrag{title}[c][c]{\Large Failure Risk}
\psfrag{zlabel}[b][t]{\Large $\lambda^*_F/c_R$}
\resizebox{1in}{!}{\includegraphics*[115,440][250,545]{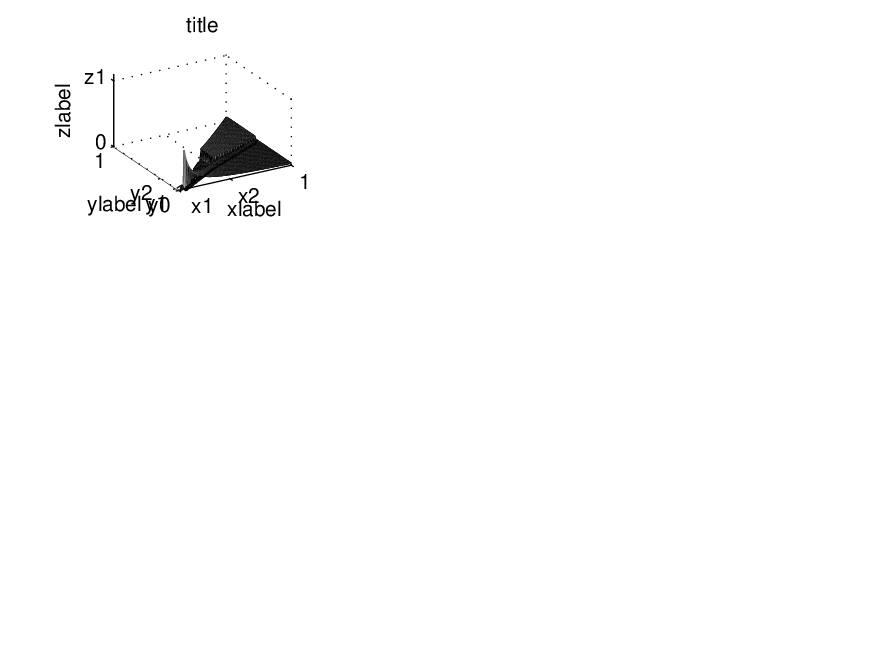}}
\\
\hspace*{-0.1in}
{\small (a)} &
\hspace*{-0.1in}
{\small (b)} &
\hspace*{-0.1in}
{\small (c)} \\
\end{tabular}
\caption{\label{fig:freeAttackPerformance} The (a) performance of the
  optimal policy in \FigRef{freeAttackPartitions}(a) and its tradeoff
  between (b)~maintenance overhead and (c)~failure risk.}
\end{figure}

\FigRef{freeAttackPartitions} summarizes the optimal partition
graphically, where we define $y_1$ as in \equRef{ruleWvsD2},
$$
y_2 = \frac{(1+p_A)p_Fp_D(2-p_R)}{p_Ap_F + p_R(p_A+p_F)},
$$
$$
y_3 = \frac{p_R(p_A+p_D) - p_F(2+p_A)(1-p_D)(1-p_R)}{(1+p_A)p_R},
$$
$$
x_1 = \frac{(1+p_A)p_F(1-p_D)(1-p_R)}{p_R(p_A+p_D) - p_F(1-p_D)(1-p_R)}
$$
and
$$
x_2 = \frac{(1+p_A)p_F(1-p_R)}{p_R(p_A+p_F)-p_F}
$$
as well as assume that the latter two quantities take positive values
less than unity.  (The portion of the design space with $c_D/c_F$
greater than $c_R/c_F$ can be ignored by virtue of the model
assumption that $c_D < c_R$.) Under strong sufficiency in
\FigRef{freeAttackPartitions}(a), the optimal policy is to
(i)~reset-under-attack $\mu^R$ when the reset cost $c_R$ is small,
(ii)~defend-under-attack $\mu^D$ when $c_R$ is large but the defend cost $c_D$
is small and (iii)~wait-under-attack $\mu^W$ when both $c_R$ and $c_D$ are
large. The three regions always meet at the point where $c_R/c_F =
x_2$ and $c_D/c_F = p_Dx_2$. It's also seen that policy $\mu^W$ cannot
be optimal if $c_R/c_F$ is below $x_2$ or $c_D/c_F$ is below $p_Dx_2$,
while policy $\mu^D$ cannot be optimal if $c_R/c_F$ is below $x_1$. In
fact, notice that, as probability $p_F$ approaches zero (i.e., as
attacks become very slow), parameters $x_2$ and $y_2$ both approach
zero at a rate inversely related to probability $p_R$, while the rate
for $y_2$ is directly related to probability $p_D$.  This implies
that, unless the defend action is highly effective or the reset action
is highly unreliable, policy $\mu^W$ is likely to be optimal---that
is, armed with a reliable reset action against slow and cost-free
attacks, a simplest reactive recovery strategy is preferable to
exercising costly defenses or initiating preemptive resets.  Also
notice that $x_1$ is strictly greater than zero if probabilities $p_D$
and $p_R$ are strictly within the unit interval.  This implies that
when both the defend and reset actions are only sometimes successful,
and even if the defend action is cost-free (i.e., even if $c_D/c_F$ is
zero), there are small enough (nonzero) values of the ratio $c_R/c_F$
to render initiating preemptive resets (i.e., policy $\mu^R$)
favorable to employing defensive countermeasures (i.e., policy
$\mu^D$). The story under weak sufficiency is similar, except that the
optimal partition involves only the inequality comparing policies
$\mu^D$ and $\mu^R$; note that $x_2$ is now negative-valued and thus
absent from \FigRef{freeAttackPartitions}(b). Finally,
\FigRef{freeAttackPerformance} plots the achieved
average-cost-per-stage $\lambda^*$ (still normalized by the reset cost
$c_R$), showing its decomposition into maintenance overhead and
failure risk as was discussed for \FigRef{relResetPerformance} in the
preceding subsection.

It is readily shown under strong sufficiency that, no matter the
probability parameters, the above quantities satisfy $x_1 < x_2$, as
sketched in \FigRef{freeAttackPartitions}(a). However, the assumption
that both $x_1$ and $x_2$ are less than unity need \emph{not} be the
case---in fact, we have $x_1 < 1$ (as well as $0 < y_3 < p_D$) if and
only if
$$
p_R >
\frac{p_F(2+p_A)(1-p_D)}{p_F(2+p_A)(1-p_D) + p_A + p_D}
$$
and $x_2 < 1$ (as well as $y_1 < p_Dx_2 < y_2 < p_D < y_3$) if and only if
$$
p_R >
\frac{p_F(2+p_A)}{p_F(2+p_A) + p_A}.
$$
Whether $x_1$ or $x_2$ exceed unity is significant because, in our
model, the cost ratio $c_R/c_F$ is itself never greater than
unity. Thus, regardless of the costs, policy $\mu^W$ can be dismissed
when given probability parameters such that $x_1 > 1$, while policy
$\mu^D$ can also be dismissed when given probability parameters such
that $x_2 > 1$. It is also readily shown that, in relation to the two
thresholds shown in \FigRef{resetSuff}, we have
$$
\frac{p_F(2+p_A)}{p_F(2+p_A) + p_A} > \frac{p_F}{p_F + p_A}
$$
and
$$
\begin{array}{l}
\displaystyle \frac{p_F(2+p_A)(1-p_D)}{p_F(2+p_A)(1-p_D) + p_A + p_D} > \\
\displaystyle \qquad \qquad \qquad \qquad \qquad \qquad
\frac{p_F(1-p_D)}{p_F(1-p_D) + p_A+p_D}
\end{array} ,
$$
while the remaining order relation depends on probabilities $p_D$ and
$p_A$; specifically, we have
$$
\frac{p_F(2+p_A)(1-p_D)}{p_F(2+p_A)(1-p_D) + p_A + p_D} >
\frac{p_F}{p_F + p_A}
$$
if and only if $p_D$ is less than $p_A/(1+p_A)$.

\begin{figure}[!h]
\centering
\psfrag{0}[c][c]{\small $0$}
\psfrag{1}[c][c]{\small $1$}
\psfrag{pR}[c][c]{reset reliability (probability $p_R$)}
\psfrag{p1}[r][r]{$\frac{p_F(1-p_D)}{p_F(1-p_D) + p_A + p_D}$}
\psfrag{p2}[r][r]{$\frac{\textcolor{white}{(}p_F}
{\textcolor{white}{(}p_F + p_A}$}
\psfrag{p3}[r][r]{$\frac{p_F(2+p_A)(1-p_D)}{p_F(2+p_A)(1-p_D) + p_A+p_D}$}
\psfrag{p4}[r][r]{$\frac{p_F(2+p_A)}{p_F(2+p_A) + p_A}$}
\psfrag{T1}[l][l]{\small \sc Insufficiency}
\psfrag{T2}[l][l]{\small \sc Weak Sufficiency}
\psfrag{T3}[l][l]{\small \sc Strong Sufficiency}
\includegraphics{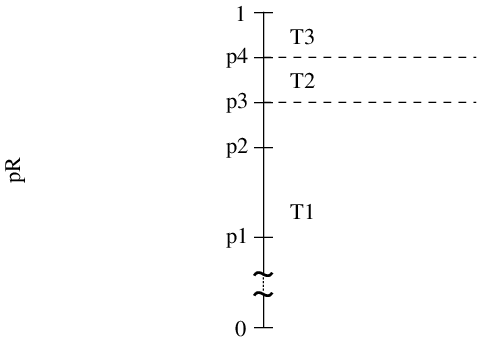} \\
(a)~If probability $p_D$ is less than $p_A/(1+p_A)$ \\[0.2in]
\includegraphics{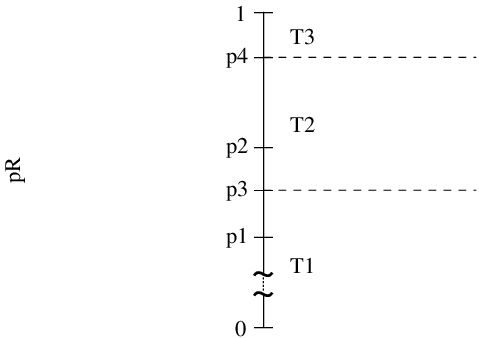} \\
(b)~If probability $p_D$ is greater than $p_A/(1+p_A)$
\caption{Refined notions of ``sufficiently reliable'' reset actions, where
comparison to \FigRef{resetSuff} depends on
the effectiveness of defend and the frequency of attack.
\label{fig:resetSuff2}}
\end{figure}

\FigRef{resetSuff2} illustrates the refined notions of ''sufficiently
reliable'' reset actions implied by the possibilities that $x_1$ or
$x_2$ exceed unity.  It may at first seem counter-intuitive that, in
comparison to \FigRef{resetSuff}, higher thresholds on probability
$p_R$ are required to achieve the different notions of sufficiency;
stated another way, the region of insufficiency in which a proactive
recovery strategy is preferable appears to be more dominant than was
the case in \FigRef{resetSuff}. Yet, in the special case that $c_A$ is
zero, or the case of no penalty associated with allowing an intrusion
attempt to proceed, intuition suggests that reactive recovery
strategies should become only more favorable, all other things equal
(i.e., the thresholds in \FigRef{resetSuff2} should be smaller than
those of \FigRef{resetSuff}). What is confounding here is that all
other things are \emph{not} equal, in the sense that the conditions
leading to the thresholds of \FigRef{resetSuff} were highly
conservative -- for ease of exposition, they only examined whether a
couple of terms in the inequalities \equRefs{CompWtoR}{CompDtoR} were
negative. The conditions leading to the thresholds of
\FigRef{resetSuff2} go several levels deeper, examining terms in
\equRefs{CompWtoR}{CompDtoR} while also accounting for the fact that,
in our model, the reset cost $c_R$ cannot exceed the failure cost
$c_F$. As a result, \FigRef{resetSuff} and \FigRef{resetSuff2} should
not be compared too directly: for example, it is easily seen that
policies $\mu^W$ and $\mu^D$ are more favorable when $c_A$ is zero,
all other things equal, by simply referring back to their penalties
$\lambda^W$ and $\lambda^D$ in \equRef{lambdaW} and \equRef{lambdaD},
respectively.

\subsection{The General Case}
This subsection builds upon the understanding gained from examining
the previous special cases to characterize inequalities
\equRefs{CompWtoR}{CompDtoR} in full generality (but, of course, still
under the idealization of perfect detection). The design space is now
in three dimensions, involving cost ratios $c_R/c_F$, $c_D/c_F$ and
$c_A/c_F$ (on the $x$-axis, $y$-axis and $z$-axis, respectively), and
each inequality now defines a planar partition of this space according
to which of the two associated policies is preferable.  In turn, all
inequalities together partition the three-dimensional volume into
components in accordance with which of the three policies is optimal.

\begin{figure}[!h]
\centering
\psfrag{xlabel}{\large $c_R/c_F$}
\psfrag{ylabel}{\large $c_D/c_F$}
\psfrag{zlabel}{\large $c_A/c_F$}
\psfrag{0}{$0$}
\psfrag{y0}{$0$}
\psfrag{1}{$1$}
\psfrag{x1}{$x_1$}
\psfrag{x2}{$x_2$}
\psfrag{y2}{$p_Dx_2$}
\psfrag{y3}{$p_D$}
\psfrag{y4}{$y_2$}
\psfrag{y5}{$y_3$}
\psfrag{z1}{$z_1$}
\psfrag{z2}{$z_2$}
\psfrag{choosew}[c][c]{choose $\mu^W$}
\psfrag{choosed}{choose $\mu^D$}
\psfrag{chooser}{choose $\mu^R$}
\begin{tabular}{cc}
\hspace*{-0.15in}
\resizebox{1.55in}{!}{\includegraphics*[20,165][205,295]
{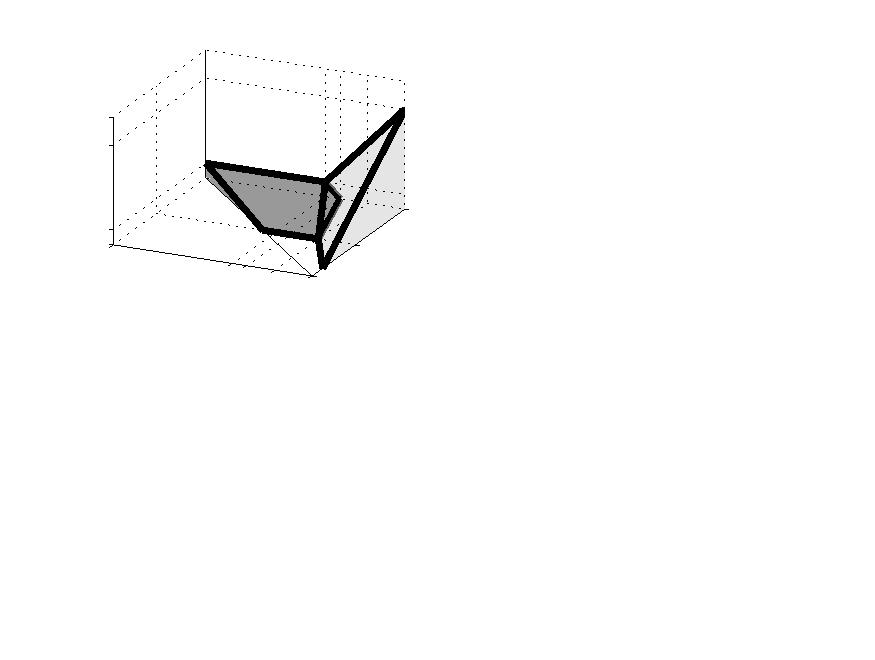}} &
\psfrag{choosed}{$\:\:\:\:$ choose $\mu^D$}
\resizebox{1.55in}{!}{\includegraphics*[20,165][205,295]
{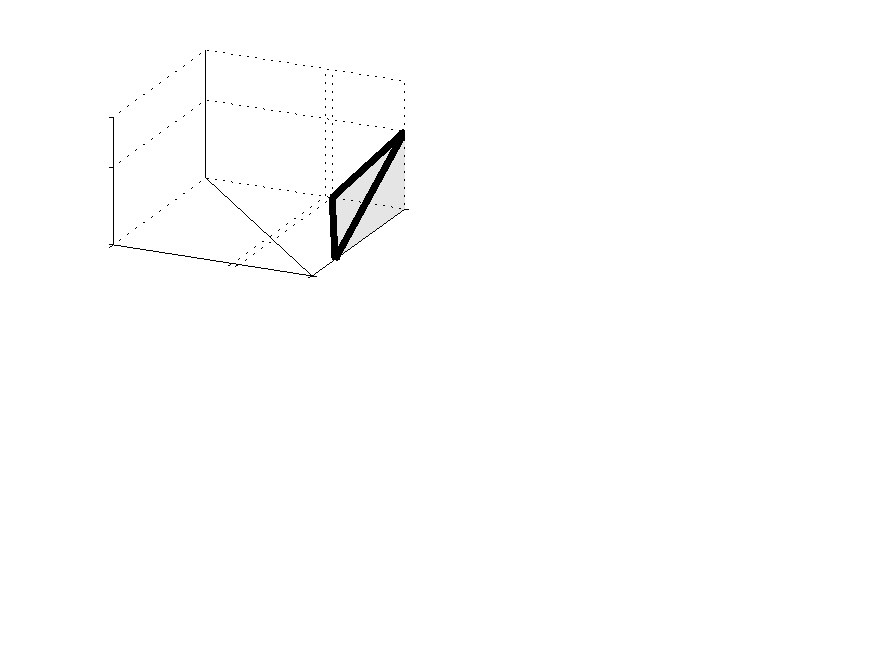}} \\
\hspace*{-0.15in}
{\small (a)} & {\small (b)}
\end{tabular}
\caption{\label{fig:generalPartitions} The partitions
  characterizing the optimal policy for the MDP model in
  \FigRef{threeStateModel} under (a)~strong sufficiency or
  (b)~weak sufficiency (as defined in \FigRef{resetSuff2}).}
\end{figure}

\FigRef{generalPartitions} summarizes the optimal partition graphically,
where we define $x_1$, $x_2$, $y_2$ and $y_3$ as in
\FigRef{freeAttackPartitions},
$$
z_1 = \frac{p_Ap_R - p_F(2+p_A)(1-p_R)}{(1+p_A)p_R}
$$
and
$$
z_2 = \frac{p_R(p_A+p_D) - p_F(2+p_A)(1-p_D)(1-p_R)}{(1+p_A)(1-p_D)p_R}
$$
as well as assume that the latter quantity takes a positive value less
than unity. Under strong sufficiency, the three planes always
intersect at two points, specifically $(x_2, p_Dx_2,0)$ and $(1,
p_D,z_1)$. The optimal policy is to (i)~wait-under-attack $\mu^W$
when, relative to cost $c_F$, both $c_R$ and $c_D$ are high but $c_A$
is low (the dark-shaded volume), (ii)~defend-under-attack $\mu^D$ when
$c_R$ is high but both $c_D$ and $c_A$ are low (the light-shaded
volume) and (iii)~reset-under-attack when $c_R$ is low or $c_A$ is
high (the unshaded volume). The story under weak sufficiency is
similar, except involving only the boundary defined by inequality
\equRef{CompDtoR}.

Notice that the partition of \FigRef{freeAttackPartitions} corresponds
to the $xy$-slice with ratio $c_A/c_F$ equal to zero, while the
partition of \FigRef{relResetPartitions} corresponds to the $zy$-slice
with ratio $c_R/c_F$ and probability $p_R$ both equal to unity.
The previous subsections on these special cases discussed in detail
how the probability parameters affect the shape of the optimal
partition and, in turn, how the cost parameters affect the choice of
the optimal policy. These discussions, suitably generalized, continue
to be relevant here e.g., consider \FigRef{generalPartitions} in the
limiting case of a perfect defend action, or when probability $p_D$
approaches unity. Observe that parameters $x_1$, $y_3$ and $z_2$
approach zero, unity and infinity, respectively, implying that the
boundary between the unshaded and lightly-shaded regions is
approaching the $x=y$ plane; that is, the region for which policy
$\mu^R$ is optimal is being entirely replaced by the region for which
policy $\mu^D$ is optimal. Meanwhile, observe that parameter $z_1$ is
constant while parameter $y_2$ approaches a value strictly less than
unity (under strong sufficiency), implying that the darkly-shaded
region pertaining to policy $\mu^W$, though certainly diminished, is
\emph{not} entirely replaced in the limit of a perfect defend action.
In the limiting case of a useless defend action, where probability
$p_D$ is near zero, we have $x_1$ near $x_2$ and $z_1$ near $z_2$ so
that the lightly-shaded volume corresponding to policy $\mu^D$
essentially disappears.

There is one aspect of \FigRef{generalPartitions} that was not
apparent in either of the special cases discussed in previous
subsections, which relates to whether parameter $z_2$ is in fact below
or above unity. In particular, it is always true as sketched that $z_1
< 1$ and $z_1 < z_2$ (and, under strong sufficiency, that $z_1 > 0$),
but the assumption that $z_2 < 1$ need \emph{not} be the case. This is
significant because, in our model, the cost ratio $c_A/c_F$ is always
less than unity. This implies that, if the probability parameters are
such that $z_2 < 1$ and if cost $c_A$ is such that $c_A/c_F > z_2$,
then policy $\mu^R$ is optimal regardless of the other costs. On the
other hand, if $z_2$ is greater than unity, then even if $c_A$ is
equal to its upper bound of $c_F$, it remains possible (e.g., if
$c_R/c_F$ is above $x_2$ and $c_D/c_F$ is below $p_Dx_2$) for policy
$\mu^D$ to be optimal.  Now, it is readily shown that $z_2 > 1$ if and
only if
$$
p_R > \frac{p_F(2+p_A)(1-p_D)}{p_F(2+p_A)(1-p_D) + \left[ p_D(2+p_A)-1\right] },
$$
where the right-hand-side is always greater than the threshold of weak
sufficiency in \FigRef{resetSuff2} but its relation to the strong
sufficiency region depends on probability $p_D$; specifically, we have
that
$$
\begin{array}{l}
  \displaystyle \frac{p_F(2+p_A)(1-p_D)}{p_F(2+p_A)(1-p_D) +
\left[ p_D(2+p_A)-1\right] } > \\
  \displaystyle \qquad \qquad \qquad \qquad \qquad \qquad \qquad \qquad
  \frac{p_F(2+p_A)}{p_F(2+p_A) + p_A}
\end{array}
$$
if and only if $p_D$ is less than $1/2$ as well as
$$
\frac{p_F(2+p_A)(1-p_D)}{p_F(2+p_A)(1-p_D) + \left[ p_D(2+p_A)-1\right] } > 1
$$
if and only if $p_D$ is less than $1/(2+p_A)$.

To summarize the preceding paragraph, in the case of
``lowly-effective'' defend actions in the sense that $p_D <
1/(2+p_A)$, parameter $z_2$ is always less than unity; in the case of
``nominally-effective'' defend actions in the sense that $1/(2+p_A) <
p_D < 1/2$, parameter $z_2$ is always less than unity under weak
sufficiency and transitions above unity within the strong sufficiency
region of \FigRef{resetSuff2}; in the case of ``highly-effective''
defend actions in the sense that $p_D > 1/2$, parameter $z_2$ is
always greater than unity under strong sufficiency and transitions
below unity within the weak sufficiency region of \FigRef{resetSuff2}.
Altogether, it follows that even in cases where an ongoing attack may
be viewed nearly as costly as a security failure (i.e., where $c_A$
approaches $c_F$), the choice between reactive or proactive recovery
still depends on the probability parameters and, in general, also the
other cost parameters; however, if armed with (i)~only a
``lowly-effective'' defend action or (ii)~only a
``nominally-effective'' defend action alongside a
``weakly-sufficient'' reset action, then a proactive recovery strategy
is best as $c_A$ approaches $c_F$ regardless of the other costs $c_D$
and $c_R$.

\section{Extensions and Related Work \label{sec:ModExt}}
The simplest model presented in \SecRef{MainModel} and analyzed in
\SecRef{PerDet} can be extended in many ways. This section suggests a
number of such extensions, also discussing the implications on the
associated analyses and providing relevant references. The discussion
specifically looks towards intrusion tolerance applications with
(i)~imperfect or controllable detectors, (ii)~multiple types of
attacks, (iii)~continuous-time dynamics or (iv)~strategic attackers.

\subsection{Imperfect or Controllable Detectors \label{ssec:ImpDet}}
The analysis of \SecRef{PerDet} assumes that each control decision has
access to perfect state information: the detectors never generate a
false-positive nor a false-negative (i.e., never indicate state $A$ if
state $N$ is true nor state $N$ if state $A$ is true, respectively)
and, similarly, the presence or absence of a security failure is never
in question (i.e., state $F$ is indicated if and only if state $F$ is
true). To extend the analysis to the realistic case of imperfect state
information, the more general formalism of a \emph{Partially
  Observable} Markov Decision Process (POMDP) \cite{KLC98:PlAct} is
required. In addition to parameters of the MDP, a POMDP model assumes
that each observation takes one of a finite set of values and then
includes probability parameters that relate to the underlying sensor
uncertainty. Moreover, much like the state transition probabilities,
these sensor observation probabilities may also depend on the
most-recently selected control action.

In our simplest three-state model, for instance, imperfect detection
of the attack state $A$ could be described by the additional
parameters of a false-positive probability $q_{A|N}$ and a
false-negative probability $q_{N|A}$, the ideal case recovered when
these probabilities are both zero. (See \cite{Kre10:MDPIT} for a preliminary
simulation-based analysis of this particular POMDP under the reliable
reset assumption, focusing on the extent to which increasing sensor
uncertainty monotonically degrades achievable performance and the loss
from optimum performance of two popular rule-based policies for
response selection.) Incorporating an additional sensing service
(e.g., a disk scan) into a defensive countermeasure is captured
by associating smaller error probabilities $q^\prime_{N|A} < q_{N|A}$
or $q^\prime_{A|N} < q_{A|N}$ to each observation that follows the
application of control $D$. A similar addition of probability
parameters can capture errors about the presence or absence of a
security failure. With multiple detectors in play,
the aggregate observation model will also include probabilities to
describe (perhaps control-dependent) correlations among the
different sensor outputs.

Regardless of the details of the observation model, the advent of
imperfect state information has dramatic ramifications on the structure
of the optimal policy and, in turn, the complexity of its analysis and
computation. This stems from the fact that, in each stage $k =
0,1,\ldots$, it is no longer true that state $x_k$ is revealed before
the control decision $u_k$ is made; rather, only an observation $z_k$
is revealed and it must be properly interpreted within the context
established from \emph{all} previously revealed information i.e., the
observation history $\allZ_{k-1} = (z_0,z_1,\ldots, z_{k-1})$ as well
as the control history $\allU_{k-1} = (u_0, u_1, \ldots,
u_{k-1})$. Under certain technical conditions (satisfied by POMDPs),
this interpretation is provided by the conditional probability
distribution $P(x_k|\allZ_k,\allU_{k-1})$, or the \emph{belief
  state}---that is, for the purposes of optimal control, the belief
state is known to be a \emph{sufficient statistic}, essentially a
lossless compression of the entire history of previously revealed
information \cite{Ber95:DPOC1}.

A POMDP model features properties that greatly simplify the
computation of successive belief states.  Firstly, because the
(hidden) state takes only finitely-many values, say $n$, the
distribution $P(x_k|\allZ_k,\allU_{k-1})$ is expressed by a length-$n$
probability vector $b_k \in [0,1]^n$, whose components sum to unity
with the $i$th component, $b_k[i]$, denoting the probability that
state $x_k$ is of its $i$th value.  Secondly, by virtue of the Markov
properties, implying in each stage $k$ that (i)~every state transition
probability satisfies
$$
p_{x_k|x_{k-1}}(u_{k-1}) = P(x_k|x_{k-1},\allZ_{k-1},\allU_{k-1})
$$
and (ii)~every sensor observation probability satisfies
$$
q_{z_k|x_k}(u_{k-1}) = P(z_k|x_{k},x_{k-1},\allZ_{k-1},\allU_{k-1}),
$$
the sequence of belief states can be computed recursively
\cite{Ber95:DPOC1}: given the preceding belief vector $b_{k-1}$ and with the
most recent control $u_{k-1}$ and observation $z_k$ both revealed, compute
\begin{equation}
\label{equ:recEst}
\bar{b}[i] := q_{z_k|i}(u_{k-1}) \sum_{j=1}^n p_{i|j}(u_{k-1}) b_{k-1}[j]
\end{equation}
for every $i = 1,2,\ldots n$ and then normalize the vector $\bar{b}$
i.e., then compute $b_k[i] := \bar{b}[i] / \left( \sum_{j=1}^n
  \bar{b}[j]\right)$ for each $i$.  The initial belief vector $b_0$ is
similarly computed: given initial state distribution $P(x_0)$ and with
observation $z_0$ revealed, normalize the vector
$$
\bar{b}[i] := q_{z_0|i}P(x_0 = i), \quad i = 1,2,\ldots, n.
$$

The question that remains, of course, is how to design the policy that in
each stage $k$ maps a belief vector $b_k$ into the control $u_k$. Decades of
theoretical research has revealed interesting structure in the optimal
policy, but also characterized the (worst~case) complexity of its
computation to be prohibitive \cite{PaT87:CoMDP}. On the upside, a number of
approximate solution algorithms have been developed that suitably
exploit the problem structure and exhibit acceptable (average-case)
complexity \cite{Abe03:ReSur}. (The solution utilized in the
simulation-based analysis of \cite{Kre10:MDPIT}, for example, is Cassandra's
publicly available implementation of the iterative ``witness''
algorithm; see http://www.pomdp.org.)  It is similarly more
challenging, compared to the MDP counterpart, to evaluate the
average-cost-per-stage performance of any given policy, which is
typically accomplished via Monte-Carlo simulation methods
\cite{DFG01:SeqMC}.

\subsection{Multiple Types of Attacks}
The model of \SecRef{MainModel} captures attacks at the highest level
of abstraction, including only two non-normal states $A$ and $F$
pertaining to whether an intrusion is ongoing or complete,
respectively.  The associated controls, defend $D$ and reset $R$,
respectively, are formulated at a comparable level of
abstraction. While different types of attacks and responses can be
represented to some extent by varying probability parameters (e.g., slow
versus fast attacks by varying probability $p_F$, reliable versus
unreliable reset by varying probability $p_R$), any single instance of
our simplest model can not capture possibilities that any particular attack
is one of many types nor that any particular response may exhibit different
degrees of effectiveness across these types. It is clear that
different types of attacks will merit different control policies, but
proper adaptation of the policy can only be accomplished if the underlying
model affords the ability to discriminate between the attack types.
Even within each individual type of attack, similar performance gains
can be realized by an ability to track finer-grained intrusion steps
or to diagnose more-detailed failure modes that merit,
respectively, different defensive countermeasures or alternative
recovery sequences

The (PO)MDP formalism extends to these more detailed representations
of attacks or failures, and the associated alternative responses, by
the introduction of additional system states and control actions. For
example, the attack state $A$ in our simplest model could be refined
into a series of states $A_1,A_2,\ldots $, each in correspondence with
a step in the intrusion within which a different defensive
countermeasure $D_1,D_2,\ldots$ is appropriate.
This modeling approach has been successfully applied to the problem of
defending a web-server against automated Internet worm attack
\cite{KrF04:aLADS}, with states pertaining to the ``bypass,'' ``download''
and ''install'' steps of a worm and with controls including
``kill-process'' and ``block-port.'' Similarly, the failure state $F$
in our model could be refined into a set of states $F_1,F_2,\ldots$,
each a different failure mode of the system for which a different
recovery action $R_1, R_2, \ldots$ is appropriate. This modeling
approach has been successfully applied to the problem of maintaining a
three-tier enterprise messaging platform in the face of imprecise fault
diagnosis \cite{JSH06:AutRe}, with states pertaining to ``component-crash'' and
``host-crash'' and with controls including ``restart-component'' and
``reboot-host.''  Other examples of computer security models
compatible with the formalism of MDPs appear in \cite{JoO97:QuMod,MGV02:MetMQ,ASH05:RTIDS}.

The introduction of additional states and controls can also capture the
possibility of multiple types of attack, each represented by its
own series of attack steps and set of failure modes.
In such models, the transition probabilities out of the normal state
$N$ and into the first step of each type of attack reflect the relative
frequencies they are
encountered (akin to probability $p_A$ in our model) e.g., probability
$p_{A_1|N}$ is greater than probability
$p_{B_1|N}$ if attack type $A$ is more common than attack type $B$.
The aggregate MDP will also depend on whether different types of
attacks can occur simultaneously, in which case there will be states
corresponding to the cross-product of those defined for each individual
attack type. Regardless of the granularity of states, the transition
probabilities serve to express different controllability properties
of each attack type i.e., the effectiveness of different defensive
countermeasures and the reliability of alternative recovery options
(akin to probabilities $p_D$ and $p_R$, respectively, in our model).
Similarly, the observation probabilities capture potentially different
observability properties of each attack type e.g., whether an
available detector is essentially blind to certain steps (or even an
entire type) of attack.

Of course, adding states and controls to an MDP model not only
increases the number of parameters to be specified, but also increases
the complexity of policy optimization and analysis. In general, given
$n$ distinct states and $m$ distinct controls, an MDP model consists
of $mn(n-1)$ transition probabilities and $mn^2$ transition costs; in the
case of imperfect state information, given $o$ distinct observations,
there are also $mn(o-1)$ observation probabilities. It is common
to store the model parameters as a set of $3m$ matrices, or three
matrices per control action $u$: an $n$-by-$n$
matrix $\mb{G}(u)$ with transition cost $g(i,u,j)$ in row $i$ and column $j$, an
$n$-by-$n$ matrix $\mb{F}(u)$ with transition probability $p_{j|i}(u)$ in row $i$ and
column $j$, and an $n$-by-$o$ matrix $\mb{H}(u)$ with observation probability $q_{z|i}(u)$
in row $i$ and column $z$. The latter two matrices, having
elements that lie in the unit interval and rows that sum to unity,
are called \emph{stochastic matrices} and feature many interesting properties
\cite{Gal96:DisSP}.
Even so, for larger models, an analysis analogous to that carried out in
\SecRef{PerDet} would yield optimal partitions in higher dimensions and, in
the case of imperfect detectors, present a larger parameter space over which
numerical methods must be employed. Such
concerns are attenuated for models in which many of the
probability/cost parameters are zero, which is fortunately often the
case in practical applications e.g., each attack state typically
permits transition to only a small subset of the full state
space.  Exploiting this model sparsity for the purpose of compact
representation, efficient computation and tractable analysis connects
to the rich fields of probabilistic graphical models \cite{Pea88:BelNw,Lau96:GrMod}
and approximate dynamic programming \cite{DeW19:PlanC,BeT96:NeuDP}.

\subsection{Continuous-Time Dynamics}
The models discussed thus far assume that the state of the system
evolves in discrete stages, neglecting the possibility that the time
duration between successive stages can itself be informative to
the decision process. Consider, for example, a staging scheme in which
control decisions are made in an event-driven manner based on reports
from the detector(s) (e.g., act upon every tenth report) and a type
of attack in which certain steps typically cause the reports to be
burstier than normal e.g., \cite{KrF04:aLADS}. As another example, historical
data from which certain attack characteristics (frequency, speed, etc.)
can be estimated may be in terms of absolute time (e.g., on-average, attack
type $A$ is attempted every half-hour, attack type $B$ is attempted every
five minutes, etc.). Endowing an automatic controller with a notion of absolute
time may also facilitate human-computer interaction
with security administrators. For example, commanded recovery sequences are
often generated via a real-time scheduling algorithm whose parameters
include time bounds on the involved operations
\cite{SNV06:ProRe,SBC07:ResIT,NgS09:QuaTu,NgS10:ReaSR}.

Generalizing the MDP formalism to account for continuous-time dynamics leads to
the so-called \emph{semi-Markov} decision process \cite{Gal96:DisSP,Ber95:DPOC2}, which assumes 
that the time durations between successive state transitions are themselves random.
One way to understand a semi-Markov process is as follows: as in an ordinary
Markov chain, it begins at time $t_0 = 0$ with a realization of state $x_0$ according
to a given initial state distribution $P(x_0)$; then, control $u_0$ is selected and
the subsequent state $x_1$ is realized according to specified transition probabilities
$p_{x_1|x_0}(u_0)$; next, time duration $\tau_1$ is realized according to a specified
distribution $P(\tau_1 \leq \tau \mid x_0, u_0, x_1)$ conditioned on the realizations
of $x_0$, $u_0$
and $x_1$, implying that state $x_1$ is entered at time $t_1 = \tau_1$ with an accrued
cost of $\tau_1g(x_0,u_0,x_1)$; the next control $u_1$ as well as the subsequent state
$x_2$ and duration $\tau_2$ are realized analogously, implying state $x_2$ is entered at
time $t_2 = \tau_1 + \tau_2$ with an additional cost of $\tau_2g(x_1,u_1,x_2)$; and so
on such that, with $t_k = \sum_{s=1}^k \tau_s$ denoting the time of the $k$th state
transition, the process holds each state $x_k$ and control $u_k$ through the
time interval $[t_k,t_{k+1})$ at an additional cost of $\tau_kg(x_k,u_k,x_{k+1})$.

More precisely, given that the process enters state $x_k = i$ at
time $t_k$ and assuming that control $u_k = u$ is constant from time $t_k$
onward, the next transition into state $x_{k+1} = j$ at time
$t_{k+1} = t_k + \tau_{k+1}$ is characterized by a distribution function
$$
Q_{j|i}(\tau|u) = P(\tau_{k+1} \leq \tau, x_{k+1} = j \mid x_k = i, u_k = u )
$$
for all $\tau > 0$. Moreover, for every stage $k = 1,2,\ldots$, each such distribution
satisfies the Markov properties
$$
\begin{array}{l}
\!\!\! Q_{j|i}(\tau|u) = \\[0.05in]
\!\! P(\tau_{k+1} \leq \tau, x_{k+1} = j \mid \allX_{k-1},\allU_{k-1},\allT_k,x_k=i,u_k=u)
 \end{array}
$$
with $\allX_k$ and $\allT_k$ denoting the analogous sequences of
random variables as those denoted by $\allU_k$ (and $\allZ_k$) in \sSecRef{ImpDet}.
In other words, the random duration $\tau_k$
between the $(k-1)$th and $k$th transitions, conditioned on the involved
state transition $i \rightarrow j$ and the constant control $u_{k-1} = u$, is independent
of the states, controls and durations of all preceding stages.
These so-called \emph{transition distributions} $Q_{j|i}(\tau|u)$ of a semi-Markov process
can be viewed as analogs to the state transition probabilities
$p_{j|i}(u)$ of an ordinary Markov chain; indeed, the former implies the latter, 
$$
p_{j|i}(u) = \lim_{\tau \rightarrow \infty} Q_{j|i}(\tau|u), 
$$
which are said to comprise the \emph{embedded Markov chain} of the semi-Markov process.
The model further assumes that each distribution $Q_{j|i}(u)$ is such that the
\emph{mean transition duration} conditioned on next entering state $j$, given by
$$
d_{i,j}(u) = \int_{0}^\infty 1 - \frac{Q_{j|i}(\tau|u)}{p_{j|i}(u)} d\tau,
$$
is finite. 

The transition costs of the MDP model and, in turn, the average-cost-per-stage
in \equRef{optCrit} also experience slight alteration. In particular, the
real-valued function $g(i,u,j)$ now represents the rate at which cost is accrued
over each time interval in which (i)~both the current state $i$ and applied control
$u$ remain unchanged and (ii)~the next transition will enter state $j$. That is,
the total cost grows as a piece-wise linear function of continuous time $t$, the slope
in each interval $[t_{k-1},t_k)$ taking the value $g(x_{k-1},u_{k-1},x_k)$. It follows
that, letting $N(t)$ denote the number of stages completed by time $t$, the
continuous-time counterpart to \equRef{optCrit} is given by
$$
\lim_{T \rightarrow \infty} \left( \frac{1}{T} \sum_{k=0}^{N(T)}
\tau_{k+1}g(x_k,u_k,x_{k+1}) \right).
$$

The above continuous-time problems are not pure Markov processes because, even though
at each instant of transition $t_k$ the future of the system statistically depends
only on the current state $x_k$ (assuming a fixed state-feedback policy $u_k = \mu(x_k)$),
at other times it depends in addition on the time $t - t_k$ elapsed since the preceding
transition. Even so, the analysis of finite-state/finite-action semi-Markov processes 
under infinite-horizon optimization criteria proceeds surprisingly similar to the ordinary 
MDP counterpart.  In particular, under rather mild technical conditions as explained in 
\cite{Ber95:DPOC2} (c.f., Proposition 3.2 in Chapter 5), it is known that (i)~the optimal policy 
takes the form $\mu:\{ 1,2,\ldots n\} \rightarrow \{ 1,2,\ldots,m\}$ and (ii)~the 
average-cost-per-stage $\lambda^u$ achieved by a candidate policy $\mu$ can be derived by 
essentially the same equations used to obtain \equRefs{lambdaW}{lambdaR} in \SecRef{PerDet}. 
Moreover, these equations (as well as optimality) depend on the transition distributions 
$Q_{j|i}(u)$ only to first-order i.e., only on the (up to) $mn(n-1)$ embedded chain 
probabilities $p_{j|i}(u)$ and the (up to) $mn^2$ average transition durations $d_{i,j}(u)$.

Unfortunately, in the realistic case of imperfect detectors, the generalization to 
continuous-time dynamics is not as readily addressed. This stems from the fact that it 
is typically impossible to ensure that decision points
align exactly with state transition times, in which case the semi-Markov nature of the
state process gives rise to violations of the Markov properties belying the belief
state recursion of \equRef{recEst}. One exception is the special case that each
transition duration $\tau_k$, conditioned on the triplet $(x_{k-1},u_{k-1},x_k)$, is
characterized by an exponential distribution that is independent of the next state
$x_k$ i.e.,
$$
Q_{j|i}(\tau|u) = p_{j|i}(u)\left[ 1 - \exp\left( -\frac{\tau}{d_i(u)}\right)\right],
\quad \tau > 0
$$
with $d_i(u)$ now interpreted as the mean transition duration out of state $i$ while under
control $u$. That the state now evolves in continuous-time as a true Markov
process is a consequence of the well-known \emph{memoryless property} of the exponential
distribution \cite{Gal96:DisSP}. In turn, the belief state computation of \equRef{recEst} applies
after appropriate modification of the embedded chain probabilities $p_{j|i}(u)$
to account for these mean transition durations $d_i(u)$ relative to the actual time $\tau$ elapsed 
since the preceding calculation. Specifically, suppose we wish to update the belief state
upon receiving observation $z_k$ given that the preceding belief vector $b_{k-1}$ was computed
(and the preceding control $u_{k-1} = u$ was selected) $\tau$ time units ago; then, we may still
employ \equRef{recEst} but only after modifying the transition probability matrix to
$$
\mb{F}(u;\tau) = \sum_{\ell=0}^\infty \bar{\mb{F}}(u)^\ell 
\left[\frac{\exp\left( -\tau/\bar{d}\right)\left(\tau/\bar{d}\right)^\ell}{\ell!}\right],
$$
where $\bar{d}$ and $\bar{\mb{F}}(u)$ denote the mean transition duration and the transition probability matrix,
respectively, of the so-called \emph{uniformized process} \cite{Gal96:DisSP} i.e., let
$\bar{d} = \min_{i,u} d_i(u)$ and the elements of matrix $\bar{\mb{F}}(u)$ equal
$$
\bar{p}_{j|i}(u) = \left\{ \begin{array}{lcl}
\frac{\bar{d}}{d_i(u)} p_{j|i}(u) & , & \mbox{if $i \neq j$} \\[0.1in]
\frac{\bar{d}}{d_i(u)} p_{i|i}(u) + 1 - \frac{\bar{d}}{d_i(u)} & , & \mbox{if $i = j$}
\end{array} \right. .
$$
Of course, designing the policy that in each stage $k$ maps the resulting belief vector $b_k$ into 
the next control $u_k$ remains (at least) as complicated as was discussed for ordinary POMDPs in \sSecRef{ImpDet}.

\subsection{Strategic Attackers}
The models considered thus far assume that attackers are, in essence, 
naive and static (in the sense of stationary statistics) over the 
time horizon that any optimized defense policy will be employed. 
That is, while per-attack variation is anticipated to the extent that 
characteristics (frequency, speed, type, etc.) are modeled statistically, 
there is no representation of the possibility that successive intrusion 
attempts are in fact under the control of an intelligent entity.  
Such a \emph{strategic attacker} is likely to be motivated by objectives 
beyond just causing a security failure (e.g., espionage, data exfiltration), 
employing its own optimized control policy to select each next action 
as a function of the observed or inferred responses of the information 
system against its past actions. An advanced persistent threat, for example, 
may strategically suppress or delay attack activity when defensive 
monitoring is expecting to be most intense. From the defender's point-of-view, 
a number of new questions arise: can we discriminate
a strategic attacker from a naive one and, if so, then deduce the end objective  
to better defend against the strategic attacker's next move(s)? 

Decision problems with two opposing controllers, or agents, is the subject of game
theory and its application to network security is a very active area of
research e.g., \cite{SHK06:GaThe,AlB03:GaThe,BaM09:ExpAt,AlB10:NeSec}. Solutions are premised
upon the foundational game-theoretic concept of equilibrium strategies e.g., in place of the one-player 
notion of an optimal policy, the idea is rather that both players adopt policies whereby each player, 
subject to the other player holding its policy fixed, sees no incentive to alter its adopted policy. 
The special case of MDPs with two opposing controllers 
are called \emph{competitive MDPs} (CMDPs), where a (Nash) equilibrium 
is known to exist under the assumption of a finite number of states and actions. 
Of course, analyzing a CMDP by computing such equilibria is no less 
difficult than in the single-player counterparts. Altogether, 
extending the model and analysis of this paper to cope with strategic 
attackers presents a formidable research challenge. Solution algorithms 
for so-called ''stochastic shortest path games'' 
\cite{PaB99:SSPGa} and ''partially observable stochastic games'' 
\cite{HBZ04:POSGa} may provide appropriate starting points.

A strategic attacker in the context of an optimal stopping problem, a 
particulary well-studied stochastic control problem with imperfect state information \cite{Ber95:DPOC1},
is formulated and analyzed in \cite{BKM12:NSCGa}. The security narrative involves a network 
for which a strategic defender must classify whether the attacker is a strategic spy or a naive spammer 
based on an observed sequence of attacks on file-servers or mail-servers: the spammer's goal is 
attacking the mail-servers, while the spy's goal is attacking the file-servers as much as possible 
before detection. It is empirically shown that an equilibrium strategy is unlikely to exist when the 
defender's policy uses a fixed-length observation window (via the standard Likelihood-Ratio Test for 
binary hypothesis testing), yet an equilibrium strategy is likely to exist when the defender's policy
adapts the observation window to the observed sequence of attacks (via the well-studied  
Sequential Probability Ratio Test, first derived in the seminal work of Abraham Wald \cite{Wal47:SeqAn}). 
Later work considers different variations of this network security classification game, obtaining 
stronger theoretical results on the existence and structure of 
(pure Nash as well as Stackelberg, the latter with defender as leader) equlibrium strategies between 
defender and spy \cite{DLM12:IntCG,SoM15:SeqDG}. 

\section{Conclusion \label{sec:Conclu}}
A key challenge in the design of modern intrusion-tolerant systems concerns the
inherent tradeoffs between the total risk of intermittent security failures, 
the total overhead of reactive defensive countermeasures and the
availability loss of proactive recovery sequences. This paper examined these
tradeoffs within the well-developed formalism of Markov Decision
Processes (MDPs), which capture not only the costs of
different alternatives but also the underlying uncertainties in
attack behaviors and response outcomes. Our analysis
revealed a number of important considerations for selecting appropriate
response policies, including how the tradeoffs between security risk, 
maintenance overhead and availability loss are affected by characteristics 
(e.g., frequency, speed, etc.) of anticipated attacks. Our results also 
gave precisely-defined notions of ``sufficiently-reliable'' reset actions i.e., 
when responses intended to reinstate normal system operation upon detecting 
a security failure are adequate for reactive recovery schemes to remain 
worthwhile, as well as ''sufficiently-effective'' defend actions i.e., when 
responses intended to prevent attack completion upon detecting an intrusion 
attempt are adequate for reactive defense schemes to remain worthwhile. 
Of course, our results were derived using a small and idealized model, 
characterizing intrusion tolerance problems at a highest level of 
abstraction---the extent to which our analyses extend, and our
conclusions extrapolate, to intrusion tolerance problems of more 
realistic size and character remains to be determined by future work and, 
ultimately, validated by experimentation with prototype security controllers 
for information systems under real-world attacks.

\bibliography{../bibtex/IntrusionTolerance,../bibtex/Textbooks}
\end{document}